\documentclass[aps, pra, twocolumn, preprintnumbers, superscriptaddress, amsmath, amssymb]{revtex4-1}
\usepackage{dcolumn}
\usepackage{amsmath}
\usepackage{epsfig}
\usepackage{float}
\usepackage{graphicx}
\usepackage{epstopdf}
\usepackage[colorlinks=true,urlcolor=blue,linkcolor=blue,citecolor=blue,breaklinks=true,bookmarks=false]{hyperref}
\usepackage{amssymb}
\usepackage{subfigure}
\usepackage{booktabs}
\usepackage{appendix}
\usepackage{makecell}

\begin{document}
\title{Atmospheric Effects on Continuous-Variable Quantum Key Distribution}

\author{Shiyu Wang}
\affiliation{State Key Laboratory of Advanced Optical Communication Systems and Networks, Center of Quantum Sensing and Information Processing, Shanghai Jiao Tong University, Shanghai 200240, China}

\author{Peng Huang}
\affiliation{State Key Laboratory of Advanced Optical Communication Systems and Networks, Center of Quantum Sensing and Information Processing, Shanghai Jiao Tong University, Shanghai 200240, China}

\author{Tao Wang}
\affiliation{State Key Laboratory of Advanced Optical Communication Systems and Networks, Center of Quantum Sensing and Information Processing, Shanghai Jiao Tong University, Shanghai 200240, China}

\author{Guihua Zeng}
\affiliation{State Key Laboratory of Advanced Optical Communication Systems and Networks, Center of Quantum Sensing and Information Processing, Shanghai Jiao Tong University, Shanghai 200240, China}

\begin{abstract}
Compared to fiber continuous-variable quantum key distribution (CVQKD), atmospheric link offers the possibility of a broader geographical coverage and more flexible transmission. However, there are many negative features of the atmospheric channel that will reduce the achievable secret key rate, such as beam extinction and a variety of turbulence effects. Here we show how these factors affect performance of CVQKD, by considering our newly derived key rate formulas for fading channels, which involves detection imperfections, thus form a transmission model for CVQKD. This model can help evaluate the feasibility of experiment scheme in practical applications. We found that performance deterioration of horizontal link within the boundary layer is primarily caused by transmittance fluctuations (including beam wandering, broadening, deformation, and scintillation), while transmittance change due to pulse broadening under weak turbulence is negligible. Besides, we also found that communication interruptions can also cause a perceptible key rate reduction when the transmission distance is longer, while phase excess noise due to arrival time fluctuations requires new compensation techniques to reduce it to a negligible level. Furthermore, it is found that performing homodyne detection enables longer transmission distances, whereas heterodyne allows higher achievable key rate over short distances.
\end{abstract}
\maketitle

\section{Introduction}
Nowadays quantum key distribution (QKD) \cite{Nicolas2002} through atmospheric turbulence channel over long distance has been realized \cite{Tunick2010, Fedrizzi2009, Liao2017-1}, and satellite-to-ground discrete variable quantum key distribution (DVQKD) \cite{Nicolas2002} over 1200 km has been verified recently \cite{Liao2017-2}. However, systems using single-photon detectors suffer from background noise \cite{Miao2005}, while homodyne or heterodyne detection with bright local oscillator (LO) acting as a filter can reduce the background noise \cite{Heim2011}. Experiments measuring Stokes operators \cite{Barbosa2002,Lorenz2004,Lorenz2006,Elser2009,Heim2010,Heim2014} have shown the filtering effect of LO. Besides, quantum-limited coherent measurements between a geostationary Earth orbit satellite and a ground station has been conducted \cite{Günthner2017}. Nevertheless, there is still no complete Gaussian-modulated coherent state (GMCS) CVQKD \cite{Grosshans2002,Grosshans2003} experiments being reported. Therefore, for future experiments and applications, it is quite necessary to analyze the atmospheric effects on GMCS CVQKD.

Recently, an elliptical beam model \cite{Vasylyev2016, Vasylyev2017} considering beam wandering, broadening and deformation has been established for quantum light through the atmospheric channel. The states of entanglement-based CVQKD through fading channels have been deduced, and the secret key rate without considering detection efficiency and noise has also been calculated \cite{Usenko2012}. However, the detection efficiency and noise have significant impacts on the final achievable key rate, and the atmospheric effects on signal transmission are not only the three effects included in the elliptical beam model but also many other effects such as arrival time fluctuations, temporal pulse broadening, angle-of-arrival fluctuations and scintillation \cite{Andrews2005}. Therefore, a comprehensive transmission model and corresponding performance analysis of CVQKD in the atmosphere are necessary.

In this report we consider GMCS CV-QKD horizontal link on the surface of the earth. We deduce a new achievable secret key rate for the atmospheric channel of CVQKD scheme with imperfect homodyne and heterodyne detection. Based on the deduced key rate formula, we consider three key parameters that affect the key rate. First, the transmittance change due to beam extinction \cite{Ricklin2006} and turbulence effects (temporal pulse broadening, beam wandering, broadening, deformation, and scintillation) \cite{Andrews2001} are considered, where extinction likes the attenuation in an optical fiber. Our results demonstrate that beam wandering, broadening and deformation are the main turbulence effects affecting the achievable key rate. Second, we consider the communication interruption caused by angle-of-arrival fluctuations \cite{Andrews2005}, and we found that the interruption probability is noticeable in the case of long-distance transmission. Third, we estimate the excess noise caused by pulse arrival time fluctuations which is found to be quite large. Based on the impacts mentioned above, we conduct a performance analysis.

This paper is organized as follows. In section \ref{sec2}, we deduce the achievable secret key rate over the atmospheric channel. In section \ref{sec3}, with the result of section \ref{sec2}, we show how atmospheric effects affect the performance of GMCS CVQKD. In section \ref{sec4}, We consider all the implications mentioned in section \ref{sec3} and perform a performance analysis. Finally we come to the conclusion and discussion in section \ref{sec5}.

\section{secret key rate through atmospheric channels}
\label{sec2}
\begin{figure}[h]
\centering
\includegraphics[scale=0.55]{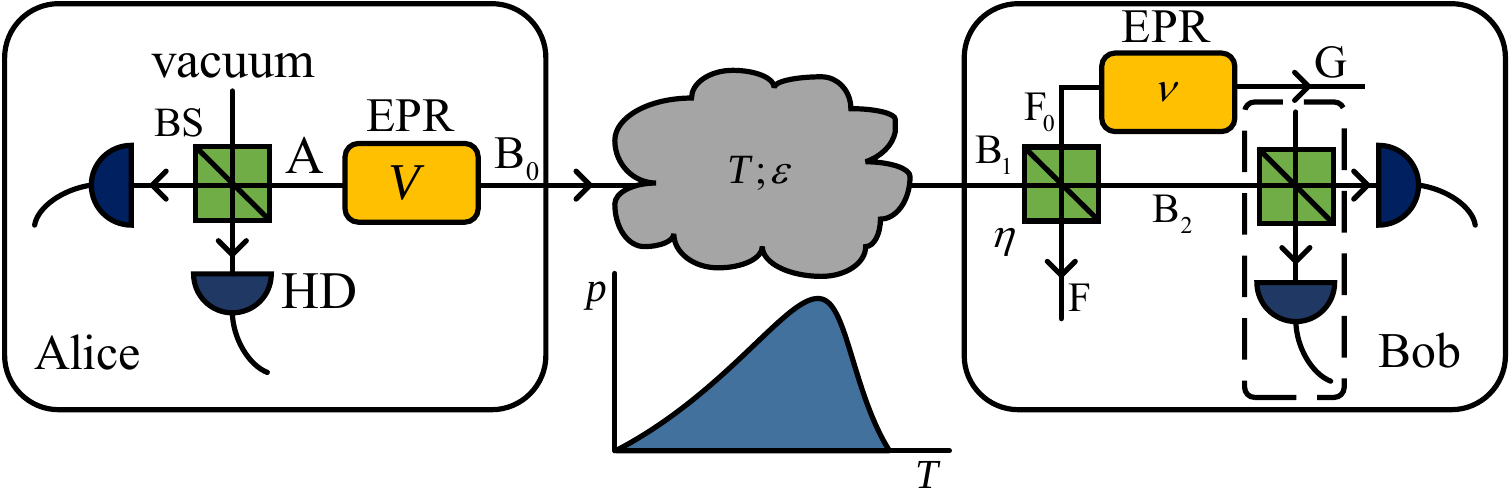}
\caption{Entanglement-based CVQKD over a fading (fluctuating) channel. HD, homodyne detection; BS, beam splitter.}
\label{SecretKeyRate}
\end{figure}

To investigate CVQKD in the atmospheric channel, we first analyze the secret key rate through fading (fluctuating) channels. The description of entanglement-based (EB) GMCS CVQKD over the fading channel is shown in FIG. \ref{SecretKeyRate}. Alice and Bob share an entangled state generated by the EPR source with variance $V$. One mode of the entangled state, ${\rm B_0}$ is transmitted to Bob through a fading channel characterized by a distribution of transmittance $T$, and Bob performs homodyne or heterodyne detection to measure field quadratures. The imperfection of the detector is described by detection efficiency $\eta$ and electronic noise $\upsilon_{el}$ contained in variance $\nu$.

In the asymptotic regime, the secret key rate $K$ is given as \cite{Weedbrook2012}
\begin{equation}
\label{eqK}
K=\beta I_{\rm AB}-\chi_{\rm BE},
\end{equation}
where $\beta$ is the reconciliation efficiency, $I_{AB}$ is the Shannon mutual information of Alice and Bob, and $\chi_{\rm BE}$ is the Holevo quantity, which can be expressed as \cite{Frossier2009}
\begin{equation}
\chi_{\rm BE}= S(\rho_{\rm E}) - \int dm_{\rm B} p(m_{\rm B}) S(\rho_{\rm E}^{m_{\rm B}}),
\end{equation}
where $m_{\rm B}$ represents the measurement of Bob, $p(m_{\rm B})$ represents the probability density of the measurement, $\rho_{\rm E}^{m_{\rm B}}$ represents the eavesdropper's state conditional on Bob's measurement result, and $S(\cdot)$ represents the Von Neumann entropy.

To calculate $I_{\rm AB}$ and $\chi_{\rm BE}$, we first need to find the covariance matrix after fluctuating channels. The covariance matrix of a two-mode squeezed vacuum state generated by the EPR source is given as
\begin{equation}
\gamma_{\rm AB}=
\begin{pmatrix}
V\mathbb{I} & \sqrt{V^{2}-1}\sigma_{z} \\ \sqrt{V^{2}-1}\sigma_{z} & V\mathbb{I}
\end{pmatrix},
\end{equation}
where $\mathbb{I}=diag(1,1)$ is the unity matrix and $\sigma_{z}=diag(1,-1)$ is the Pauli matrix. After a channel with excess noise $\varepsilon$ and random variable transmittance $T$, the covariance matrix becomes \cite{Usenko2012}
\begin{equation}
\label{eq4}
\gamma_{\rm AB_1}=
\begin{pmatrix}
V\mathbb{I} & \langle\sqrt{T}\rangle\sqrt{V^{2}-1}\sigma_{z} \\ \langle\sqrt{T}\rangle\sqrt{V^{2}-1}\sigma_{z}
& \langle T\rangle(V+1/\langle T\rangle-1+\varepsilon)\mathbb{I}
\end{pmatrix}.
\end{equation}
It can be seen from Eq.(\ref{eq4}) that the influence of the fading channel is primarily reflected in $\langle T\rangle$ and $\langle\sqrt{T}\rangle$. Thus, considering the detection efficiency $\eta$ and electronic noise $\upsilon_{el}$, we can obtain the mutual information of Alice and Bob for homodyne detection
\begin{equation}
\label{eq5}
I_{\rm AB}^{\rm hom}=\frac{1}{2}\log_2\frac{1}{1-\frac{\langle\sqrt{T}\rangle^2(V-1)}{\langle T\rangle(V+\chi_{f}^{\rm hom})}},
\end{equation}
where $\chi_{f}^{\rm hom}=(1+\upsilon_{el})/\eta\langle T\rangle-1+\varepsilon$, and for heterodyne detection
\begin{equation}
\label{IAB_het}
I_{\rm AB}^{\rm het} = \log_2\frac{1}{1-\frac{\langle\sqrt{T}\rangle^2(V-1)}{\langle T\rangle(V+\chi_{f}^{\rm het})}},
\end{equation}
where $\chi_{f}^{\rm het} = 2(1+\upsilon_{el})/\eta\langle T\rangle-1+\varepsilon$.

We can also estimate the Holevo quantity $\chi_{\rm BE}$ based on Eq.(\ref{eq4}), which can be simplified to \cite{Frossier2009}
\begin{equation}
\chi_{\rm BE} = \sum_{i=1}^{2}G(\frac{\lambda_i-1}{2}) - \sum_{i=3}^{5}G(\frac{\lambda_i-1}{2}),
\end{equation}
where $G(x)=(x+1)\log_2(x+1)-x\log_2x$. Nevertheless, the five symplectic eigenvalues in \cite{Frossier2009} can not be used directly. Thus, we deduce the symplectic eigenvalues of both homodyne and heterodyne detection for fading channels (for details see Appendix \ref{appendixA}). It is noteworthy that the results presented in Appendix \ref{appendixA} are not only applicable to the atmospheric turbulence channel but also to other channels whose transmittance changes randomly, such as underwater channels.

However, for the atmospheric channel which may cause angle-of-arrival fluctuations, Eq.(\ref{eqK}) should be revised to
\begin{equation}
\label{eq14}
K_{\rm atm} = (1-P)(\beta I_{\rm AB}-\chi_{\rm BE}),
\end{equation}
where $P$ stands for interruption probability due to angle-of-arrival fluctuations. The details of angle-of-arrival fluctuations for Eq.(\ref{eq14}) will be demonstrated in section \ref{secIP}.

\section{Atmospheric channel effects on CVQKD}
\label{sec3}
The secret key rate demonstrated in Eq.(\ref{eq14}) indicates that parameters ($T$, $\varepsilon$ and $P$) closely related to atmospheric effects should be analyzed in depth so that we can approximately assess the performance of atmospheric CVQKD by developing the method proposed in Ref. \cite{Usenko2012}. In this section, some well-developed models in atmospheric optical communications will be employed to accomplish the assessment of performance, in which we will build links between the models and atmospheric CVQKD, and show how much influence would be. Besides, a new phase excess noise caused by pulse arrival fluctuations will be derived.

Atmospheric channel effects primarily include beam extinction and turbulence effects. On the one hand, extinction is caused by absorption and scattering by molecules and aerosol which leads to attenuation of light intensity. On the other hand, random variations in the refractive index of atmosphere may cause pulse temporal broadening, transmittance fluctuations (signal fading), communication interruptions, and extra phase excess noise.
\begin{figure}[htbp]
\centering
\includegraphics[scale=0.65]{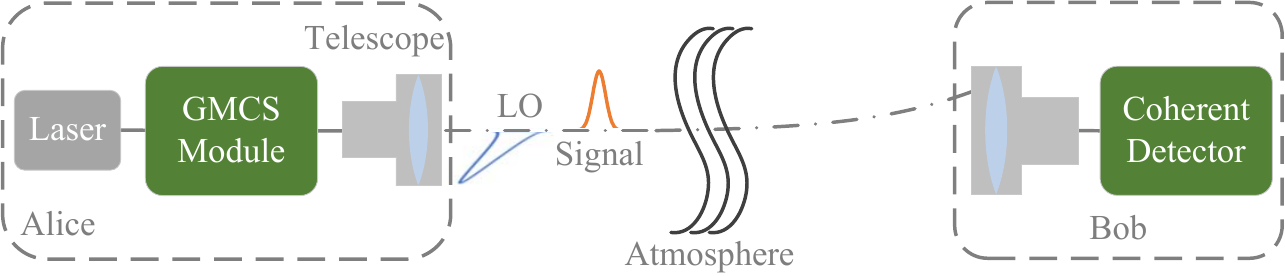}
\caption{Schematic diagram of prepare-and-measure GMCS CVQKD.}
\label{PM}
\end{figure}

In this paper, we consider that (temporal and spatial) Gaussian beam is transmitted horizontally on the surface of the Earth, as depicted in Fig. \ref{PM}. Time and polarization multiplexed LO and signal pulses are generated by the GMCS module and collimated by the telescope. Subsequently, the pulses that undergo turbulence and extinction, are collected by a telescope and measured by coherent detector, i.e., homodyne or heterodyne. It is noteworthy that LO can also be generated in Bob's terminal \cite{Qi2015,Wang2018}. This "local" LO scheme is capable of avoiding loopholes due to sending LO through the channel, but not mature relative to the scheme of simultaneous transmission of LO and signal, which has been developed over 15 years since the first experiment \cite{Grosshans2003}. In order to integrate with optical fiber systems, the wavelength of laser is chosen as 1550 nm which is also in the atmospheric transmission window. The Rytov variance is employed to describe the strength of turbulence which is given by \cite{Andrews2001}
\begin{equation}
\sigma_1^2 = 1.23C_n^2k^{7/6}L^{11/6},
\end{equation}
where $k=2\pi/\lambda$ is the optical wave number, $\lambda$ is the wavelength of light, $L$ is the horizontal propagation distance, and $C_n^2$ is the index of refraction structure parameter. Many models of $C_n^2$ have been proposed over the years \cite{Good1988}. However, since the link is assumed to be located within the boundary layer, it is rather reasonable to assume $C_n^2$ (median) to be constant as shown in Table \ref{tableCn2}, which are based on results of long-term radiosonde measurement conducted in Hefei, Anhui, China \cite{Wang2015}. Now let us estimate the impact of atmospheric effects on the performance of CVQKD.
\begin{table}[htbp]
\centering
\caption{The values (median) of $C_n^2$ within the boundary layer of four seasons.}
\begin{tabular}{ccccc}
	\toprule[0.75pt]
	& Spring & Summer & Autumn & Winter \\
	\midrule[0.5pt]
	\makecell[cc]{$C_n^2 $ \\ $\rm m^{-2/3} \times 10^{-15}$} & 2.03 & 2.12 & 5.56 & 7.46 \\
	\bottomrule[0.75pt]
\end{tabular}
\label{tableCn2}
\end{table}

\subsection{Transmittance: Beam Extinction}
For CVQKD, beam extinction means that the transmittance associated with wavelength and propagation path decreases as transmission distance increases. For horizontal paths, the transmittance is given by \cite{Ricklin2006}
\begin{equation}
\label{eqHorizontalExtinction}
T_{\rm ext}(L) = e^{-\alpha(\lambda)L},
\end{equation}
where the total extinction coefficient $\alpha(\lambda)$ comprises the aerosol scattering, aerosol absorption, molecular scattering, and molecular absorption terms:
\begin{equation}
\alpha(\lambda) = \alpha_{\rm sca}^{\rm aer}(\lambda) + \alpha_{\rm abs}^{\rm aer}(\lambda) + \alpha_{\rm sca}^{\rm mol}(\lambda) + \alpha_{\rm abs}^{\rm mol}(\lambda).
\end{equation}
There are some models that can be used to estimate the transmittance of a particular environmental conditions for CVQKD, such as LOWTRAN, MODTRAN and FASCODE \cite{Ricklin2006}. It is assumed that the horizontal link is located in the mid-latitude countryside and has a visibility of 23 km, then the extinction coefficients can be estimated by LBLRTM \cite{Wang2015}. Since $C_n^2$ in spring and summer are close to each other, here we consider the worse one, summer. Besides, the strongest turbulence occurs in winter, thus the case of winter should be considered. The extinction coefficients are listed in Table \ref{TableExt}.
\begin{table}[h]
	\centering
	\caption{The extinction coefficients in summer and winter in ${\rm km}^{-1}$.}
	\label{TableExt}
	\begin{tabular}{ccccc}
		\toprule[0.75pt]
		Seasons & $\alpha_{\rm sca}^{\rm mol}$ & $\alpha_{\rm abs}^{\rm mol}$ & $\alpha_{\rm sca}^{\rm aer}$ & $\alpha_{\rm abs}^{\rm aer}$ \\
		\midrule[0.5pt]
		Summer & $1.64\times10^{-4}$ & $3.35\times10^{-3}$ & $2.52\times10^{-2}$ & $5.49\times10^{-3}$ \\
		Winter & $1.77\times10^{-4}$ & $8.56\times10^{-4}$ & $2.52\times10^{-2}$ & $5.49\times10^{-3}$ \\
		\bottomrule[0.75pt]
	\end{tabular}
\end{table}

\subsection{Transmittance: Temporal Pulse Broadening}
\label{secTPB}
In this section we will study the transmittance change due to temporal pulse broadening under weak turbulence, regardless of the beam extinction.

Temporal pulse broadening is primarily caused by two mechanisms \cite{Liu1979}: first, the difference in arrival time of each individual pulse when only single scattering is affecting the pulse, i.e., pulse arrival time fluctuations (pulse wandering) which is also responsible for extra excess noise and will be further discussed in section \ref{secnoise}, second, the pulse spreading brought by multiple scattering process of each pulse. The combination of these two mechanisms causes temporal pulse broadening, which can be described by the averaged pulse intensity $\langle I(\textbf{r},L;t) \rangle$, or referred to as mean irradiance.

\begin{figure}[h]
\centering
\includegraphics[scale=0.7]{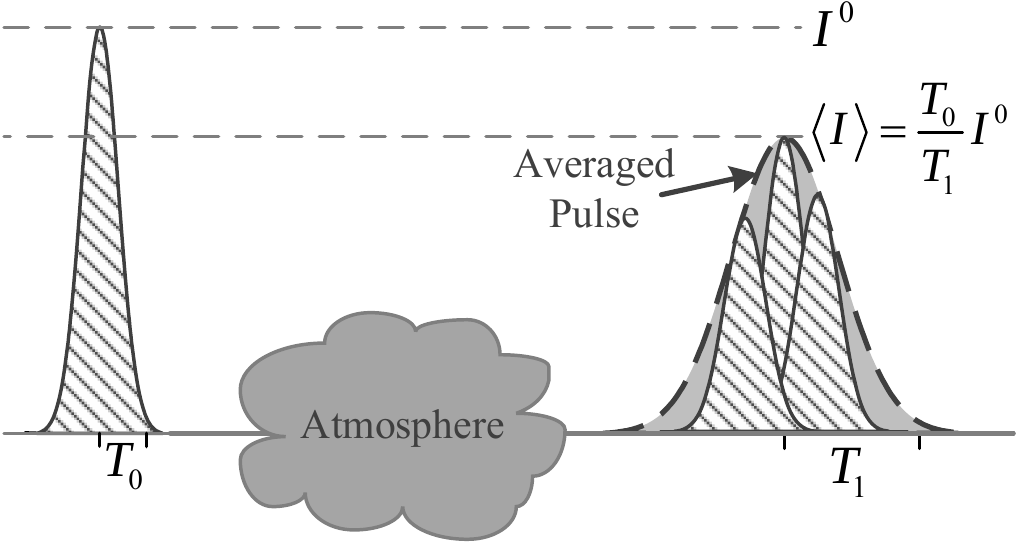}
\caption{The overall temporal pulse broadening due to pulse spreading and wandering.}
\label{BeamBroadenging}
\end{figure}

Without loss of generality, here we consider the temporal pulse broadening of Gaussian pulse \cite{Andrews2005}, whose intensity has a shape of $I(t) = \exp(-t^2/T_0^2)$, where
\begin{equation}
T_0 = \frac{R_{\rm dut}}{2f_{\rm PRF}},
\end{equation}
is the pulse half-width. Here, $R_{\rm dut}$ and $f_{\rm rep}$ are the duty ratio and pulse recurrence frequency (PRF), respectively. The temporal pulse broadening of Gaussian pulse is demonstrated in FIG. \ref{BeamBroadenging}. The temporal width of pulse is broadened by the atmosphere, and the pulse intensity is also attenuated.

The free-space irradiance of a collimated beam under the near-field ($\Omega\gg1$) and far-field ($\Omega\ll1$) approximations is given by \cite{Ziolkowski1992}
\begin{equation}
\label{freespaceNear}
I^0(\textbf{r},L;t) = \exp\left( -\frac{2r^2}{W_0^2} \right)\exp\left[ -\frac{2(t-L/c)^2}{T_0^2} \right],
\end{equation}
\begin{equation}
\label{freespaceFar}
\begin{split}
I^0(\textbf{r},L;t) = & T_0 \left( \frac{W_0^2}{2Lc} \right)^2 \frac{\omega^2T_0^4+T_0^2+\left( \frac{W_0r}{Lc} \right)^2}{[T_0^2+\left( \frac{W_0r}{Lc} \right)^2]^\frac{5}{2}} \\
& \times\exp\left[ -\frac{1}{2}\frac{(\omega T_0W_0r)^2}{(LcT_0)^2+(W_0r)^2} \right] \\
& \times\exp\left[ -\frac{2(t-\frac{L}{c}-\frac{r^2}{2Lc})^2}{T_0^2} \right],
\end{split}
\end{equation}
respectively, where $c$ is the light speed in free space, $\omega$ is the angular frequency of the light, $W_0$ is the beam-spot radius at the transmitter and $\Omega=kW_0^2/2L$ is the Fresnel parameter. However, since the amount of spreading and arrival time of each pulse are different, Eq. (\ref{freespaceNear}) and (\ref{freespaceFar}) are not able to be applied to characterize broadening in turbulence. This is exactly why $\langle I(\textbf{r},L;t) \rangle$ is required. 

Under near-field assumption the mean irradiance in weak turbulence is given by \cite{Young1998}
\begin{equation}
\label{eq54}
\langle I(\textbf{r},L;t) \rangle = \frac{T_0}{T_1}\exp\left( -\frac{2r^2}{W_0^2} \right)\exp\left[ -\frac{2(t-L/c)^2}{T_1^2} \right],
\end{equation}
where
\begin{equation}
\label{T1}
T_1 = \sqrt{T_0^2+8a_1}
\end{equation}
with $a_1 = 0.39C_n^2LL_0^{5/3}c^{-2}$, where $L_0$ is the outer scale of turbulence. The quantity $T_1$ can be considered as estimation of the broadened half-width at receiver.

\begin{figure}[h]
	\centering
	\includegraphics[scale=0.55]{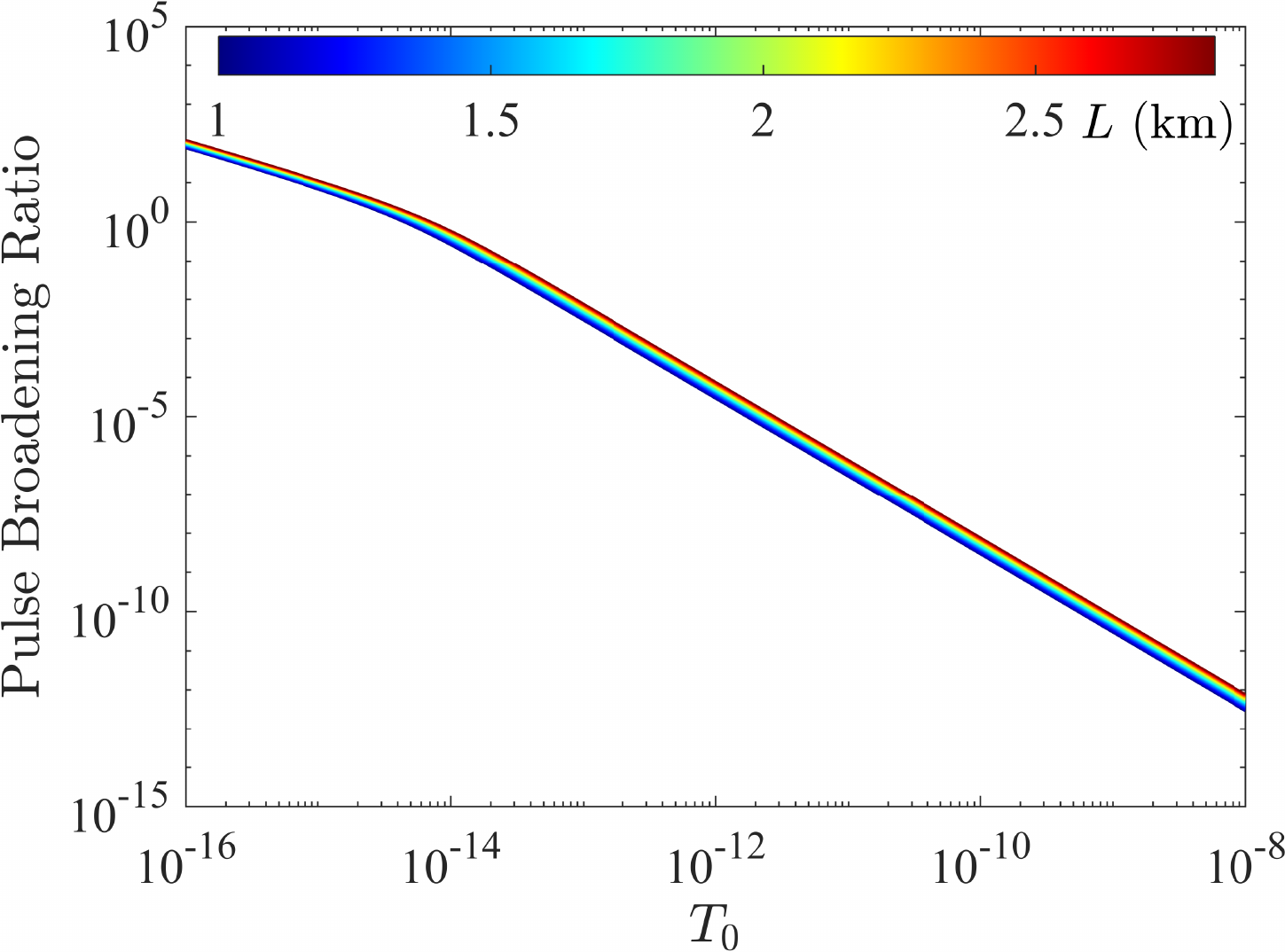}
	\caption{(Color Online) The pulse broadening ratio varies with $T_0$ at different distance in winter (from dark blue to dark red).}
	\label{PulseBroadening}
\end{figure}

Under far-field assumption the mean irradiance in weak turbulence is given as \cite{Kelly1999}
\begin{equation}
\label{eq56}
\begin{split}
\langle I(\textbf{r},L;t) \rangle = & \frac{T_0^2}{T_1} \left( \frac{W_0^2}{2Lc} \right)^2 \frac{\omega^2T_0^4+T_0^2+\left( \frac{W_0r}{Lc} \right)^2}{[T_0^2+\left( \frac{W_0r}{Lc} \right)^2]^\frac{5}{2}} \\
& \times\exp\left[ -\frac{1}{2}\frac{(\omega T_0W_0r)^2}{(LcT_0)^2+(W_0r)^2} \right] \\
& \times\exp\left[ -\frac{2(t-\frac{L}{c}-\frac{r^2}{2Lc})^2}{T_1^2} \right].
\end{split}
\end{equation}
It can be seen from Eq.(\ref{eq54})-(\ref{eq56}) that the turbulence-induced temporal pulse broadening in both near-field and far-field approximations can be characterized by $T_1$. Here, we define the pulse broadening ratio as $(T_1-T_0)/T_0$ which only in the femtosecond order will have a significant impact, as indicated by FIG. \ref{PulseBroadening}. The outer scale on the ground is assumed to be 0.4 m \cite{Lukin2005}. The result is for winter whose turbulence strength is strongest.

\begin{figure}
	\centering
	\includegraphics[scale=0.55]{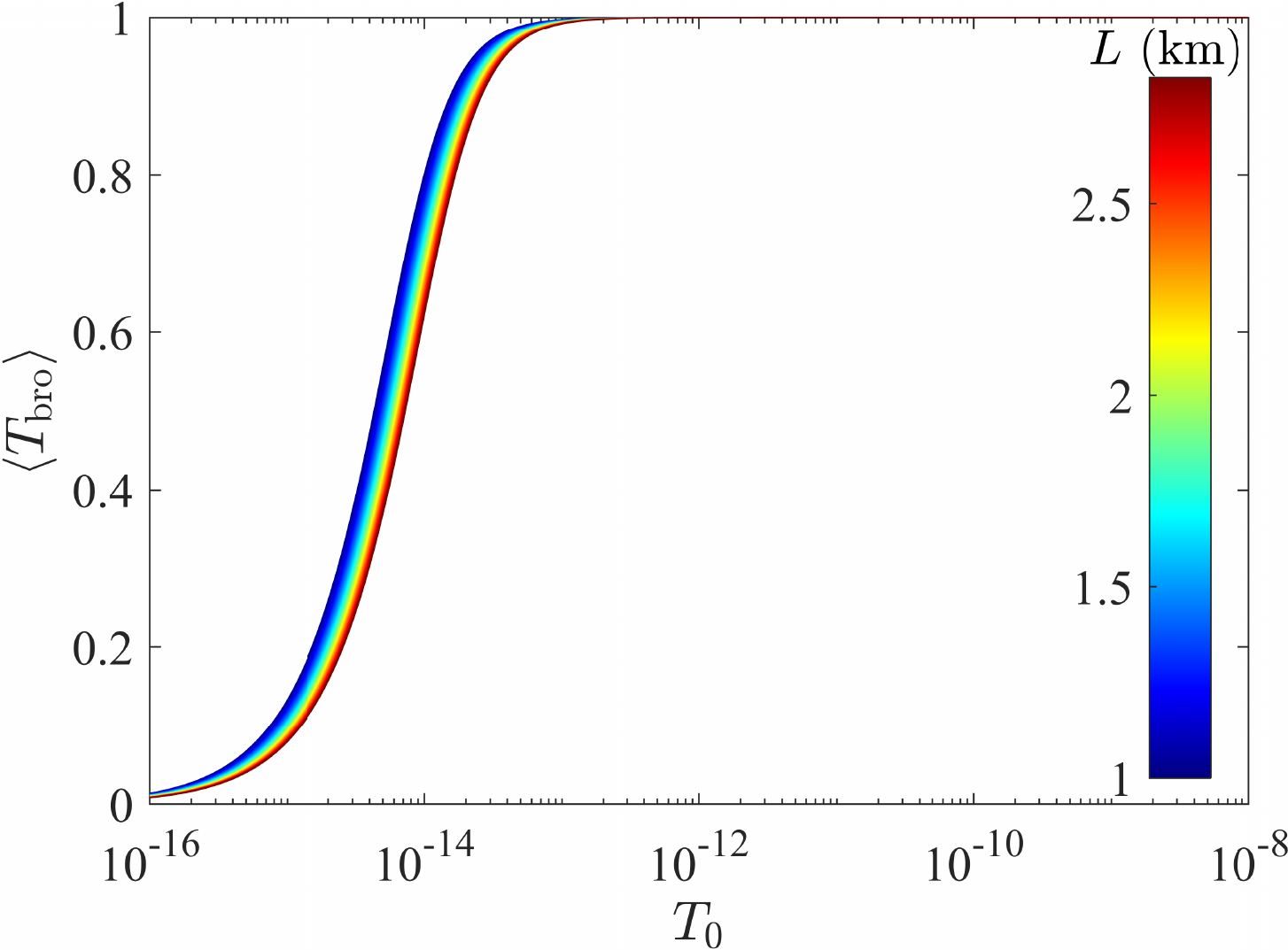}
	\caption{(Color Online) $\left\langle T_{\rm bro} \right\rangle$ varies with $T_0$ at different distance in winter (from dark bule to dark red).}
	\label{Broadening}
\end{figure}

Now we consider the transmittance change due to temporal pulse broadening. Comparing Eq.(\ref{freespaceNear}) and (\ref{eq54}) shows that pulse broadening will result in a $T_0/T_1$-fold attenuation of the average light intensity of the received signal, as shown in FIG. \ref{BeamBroadenging}. The same result can be found by comparing Eq.(\ref{freespaceFar}) with Eq.(\ref{eq56}). Thus, the mean value of pulse broadening introduced transmittance can be expressed as
\begin{equation}
\label{TransBro}
\left\langle T_{\rm bro} \right\rangle  = \frac{T_0}{T_1}.
\end{equation}
The mean transmittance in winter is demonstrated in FIG. \ref{Broadening}, $\left\langle T_{\rm bro} \right\rangle$ varies quickly from the femtosecond level to the picosecond level, but after the picosecond level, $\left\langle T_{\rm bro} \right\rangle$ is approximately equal to one. Therefore, $\left\langle \sqrt{T_{\rm bro}} \right\rangle$ is also approximately equal to one (for details see Appendix \ref{Tbro}). In other words, the transmittance introduced by pulse broadening is actually negligible in the regime of weak turbulence. However, in the regime of strong turbulence, the analysis of Eq. (\ref{TransBro}) requires a large amount of numerical calculations \cite{Chen2012}. In Ref. \cite{Chen2012}, the results also show that broadening is only perceptible on the order of femtosecond in strong turbulence, i.e., the broadening-induced transmittance approaches one, thus negligible. Therefore, there is no need to consider the transmittance change caused by pulse broadening in the following performance analysis in section \ref{sec4}.

\subsection{Transmittance: Beam wandering, broadening, deformation and scintillation}
\label{secT2}
In this subsection we will concentrate on transmittance fluctuations (signal fading) which is primarily caused by beam wandering, beam broadening, beam deformation, and scintillation.

\begin{figure}[h]
	\centering
	\includegraphics[scale=0.55]{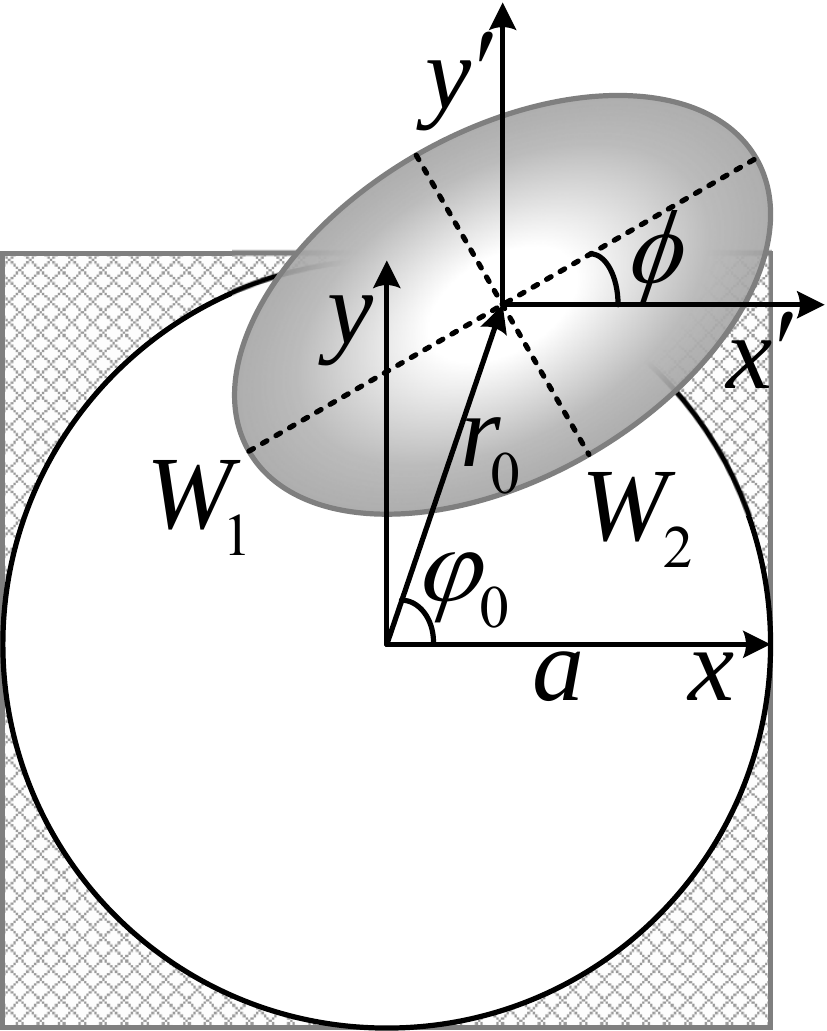}
	\caption{Elliptical beam incident on a receiving lens of radius $a$ after passing through the atmospheric channel.}
	\label{EllipticalBeam}
\end{figure}

The elliptical beam model \cite{Vasylyev2016} well describes beam wandering, broadening and deformation in weak and strong turbulence, as shown in FIG. \ref{EllipticalBeam}. However, note that the moderate-to-strong transition regime of this model has not been clarified yet, the corresponding performance analysis of CVQKD in this regime may need a better transmittance model. Weak, moderate and strong turbulence correspond to $\sigma_1^2<1$, $\sigma_1^2\approx1...10$, and $\sigma_1^2>10$, respectively. The set $\{ x_0,y_0,W_1,W_2,\phi \}$ uniquely describes the elliptic spot at the receiving aperture plane, where $(x_0, y_0)^T$ is the beam-centroid position which is equal to $(r_0\cos\varphi_0, r_0\sin\varphi_0)^2$, $W_i$, $i=1,2$, are semiaxes of the elliptic spot, and $\phi\in[0,\pi/2)$ is the angle between semiaxis $W_1$ and the $x$ axis.

Here we define $\zeta=\phi-\varphi_0$, the transmittance is then given approximately by \cite{Vasylyev2016}
\begin{equation}
\label{eq20}
T_{\rm ell}  = T_{\rm ell,0} \exp\left\lbrace -\left[ \frac{r_0/a}{R\left( \frac{2}{W_{\rm eff}(\zeta)} \right)} \right]^{Q\left(\frac{2}{W_{\rm eff}(\zeta)}\right)} \right\rbrace,
\end{equation}
where $a$ is the receiving aperture radius, and specific expressions of the other parameters are shown in Appendix \ref{ELLparameters}.

The transmittance $T_{\rm ell}$ is a function of five real parameters, $\{x_0,y_0,\Theta_1,\Theta_2,\phi\}$, where $W_i^2=W_0^2\exp\Theta_i$. In the isotropic turbulence case, $\phi$ can be seen as a uniformly distributed random variable, having no correlations with other parameters. Considering $\langle x_0\rangle=\langle y_0\rangle=0$, there are also no correlations among normally distributed $x_0$, $y_0$, and $\Theta_i$. Consequently, only $\langle \Delta x_0^2\rangle$, $\langle \Delta y_0^2\rangle$, $\langle\Delta\Theta_i^2\rangle$, $\langle\Theta_i\rangle$, and $\langle\Delta\Theta_1\Delta\Theta_2\rangle$ are required to determine the four-dimensional Gaussian random variable $\textbf{v}=(x_0, y_0, \Theta_1, \Theta_2)^T$. The mean values and covariance matrix elements are shown in TABLE \ref{tab1} (see Appendix \ref{TABLE}).

\begin{figure}[h]
	\centering
	\includegraphics[scale=0.55]{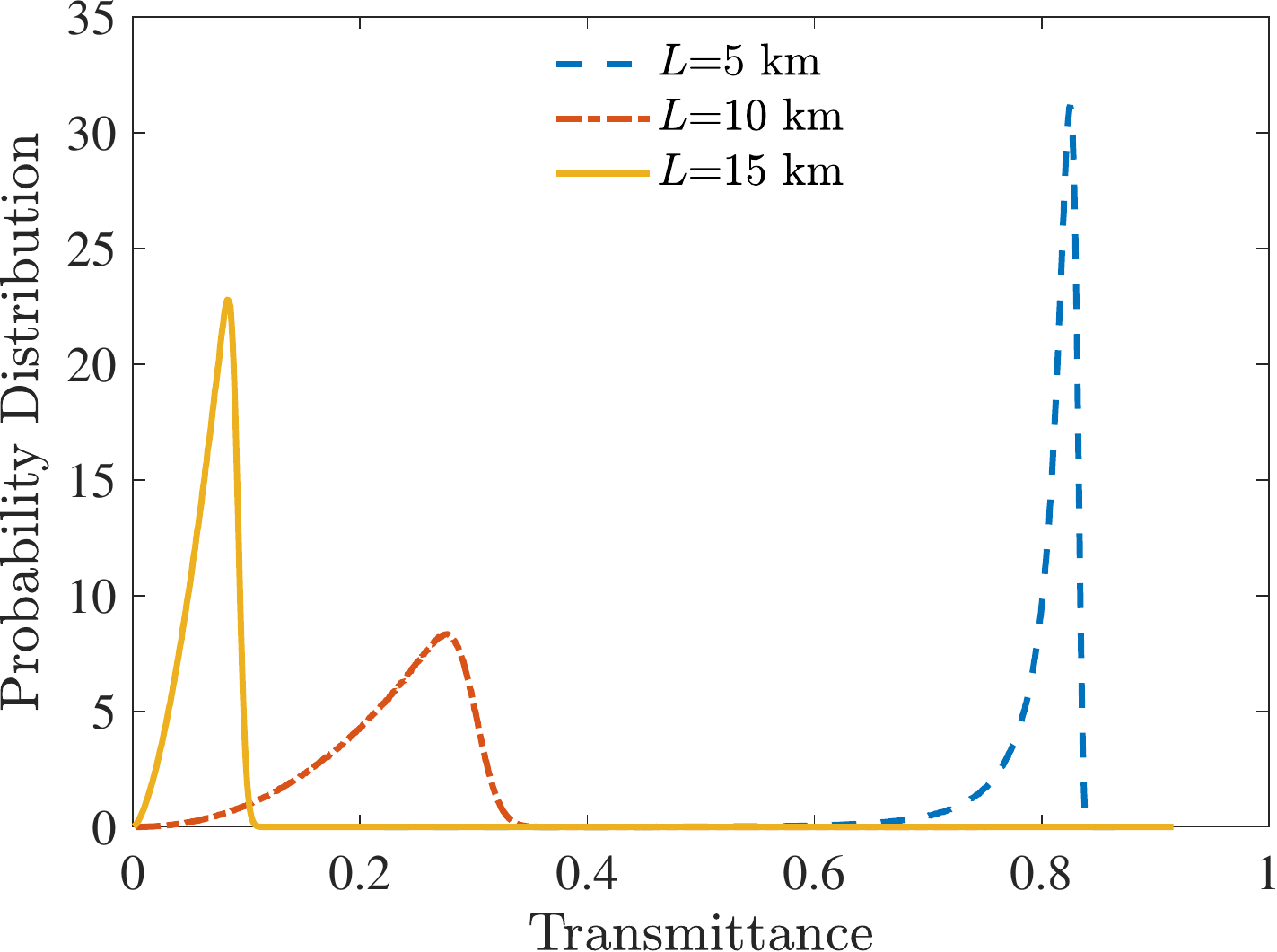}
	\caption{(Color Online) The probability density function of transmittance $T_{\rm ell}$ on the ground in summer at distance 5, 10, and 15 km. Further parameters: $a=110$ mm, $W_0=80$ mm.}
	\label{EllipticBeamPDF}
\end{figure}

Based on Eq.(\ref{eq20}), Appendix \ref{ELLparameters} and TABLE \ref{tab1}, the probability distribution of $T_{\rm ell}$ can be estimated by Monte Carlo simulations. The transmittance probability density function (PDF) of $T_{\rm ell}$ in summer on the ground is shown in FIG. \ref{EllipticBeamPDF}, with extinction involved. The receiving aperture radius $a$ and the initial beam-spot radius $W_0$ in FIG. \ref{EllipticBeamPDF} are assumed to be 110 mm and 80 mm, respectively.

Now we consider transmittance fluctuations due to scintillation which is not incorporated in the elliptical beam model. The total transmittance is considered as multiplication of Eq. (\ref{eq20}) and transmittance due to scintillations, which approximately gives a lower bound for atmospheric CVQKD, so as to make a conservative estimation of performance of CVQKD. Inherently, turbulence effects contained in the elliptical model and scintillations should be combined together, and this will be further investigated in the next step of our work.

\begin{figure}[h]
	\centering
	\subfigure[]{
		\centering
		\label{NoScintillationShematic}
		\includegraphics[scale=0.32]{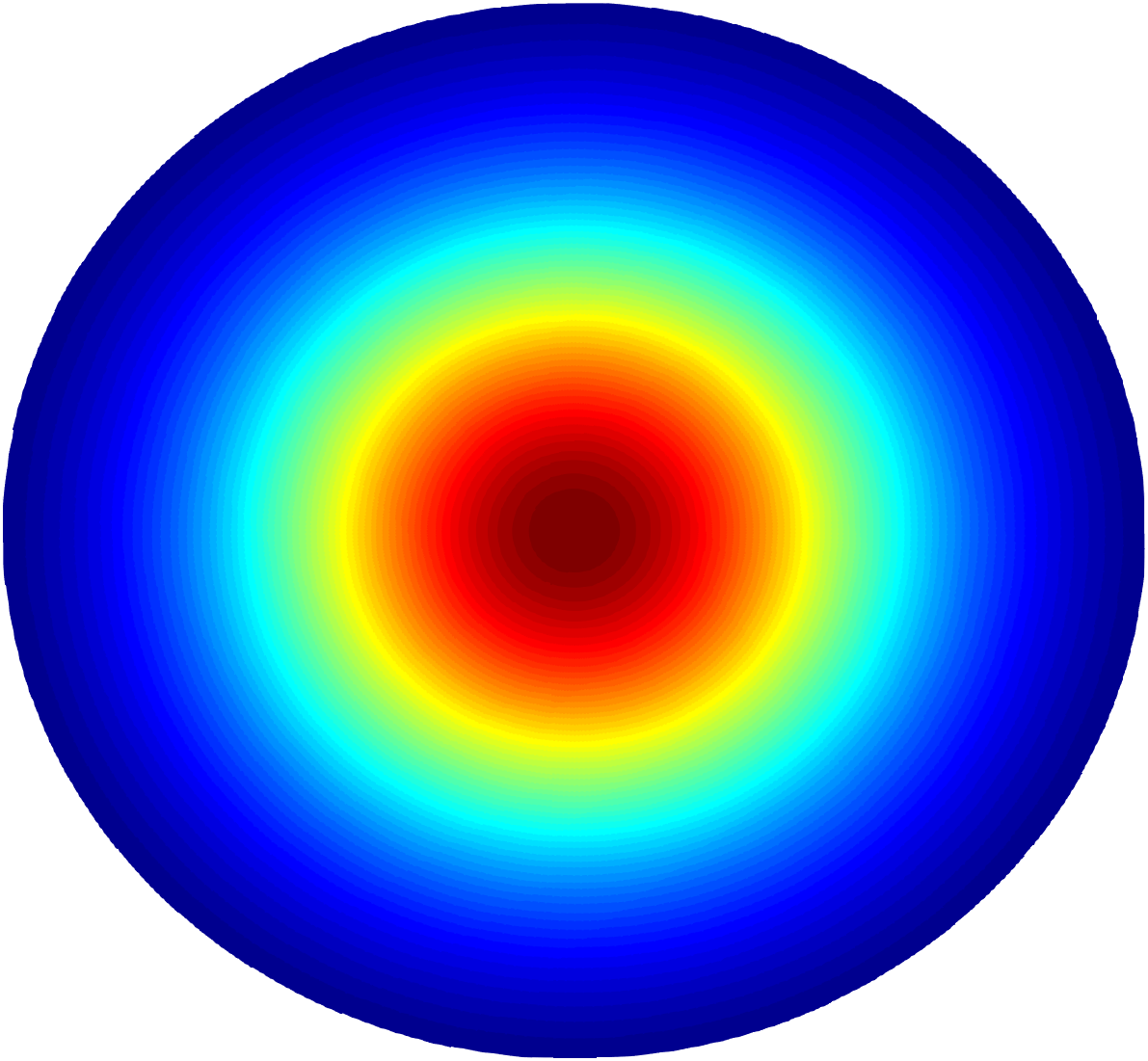}
	}
	\subfigure[]{
		\centering
		\label{ScintillationShematic}
		\includegraphics[scale=0.32]{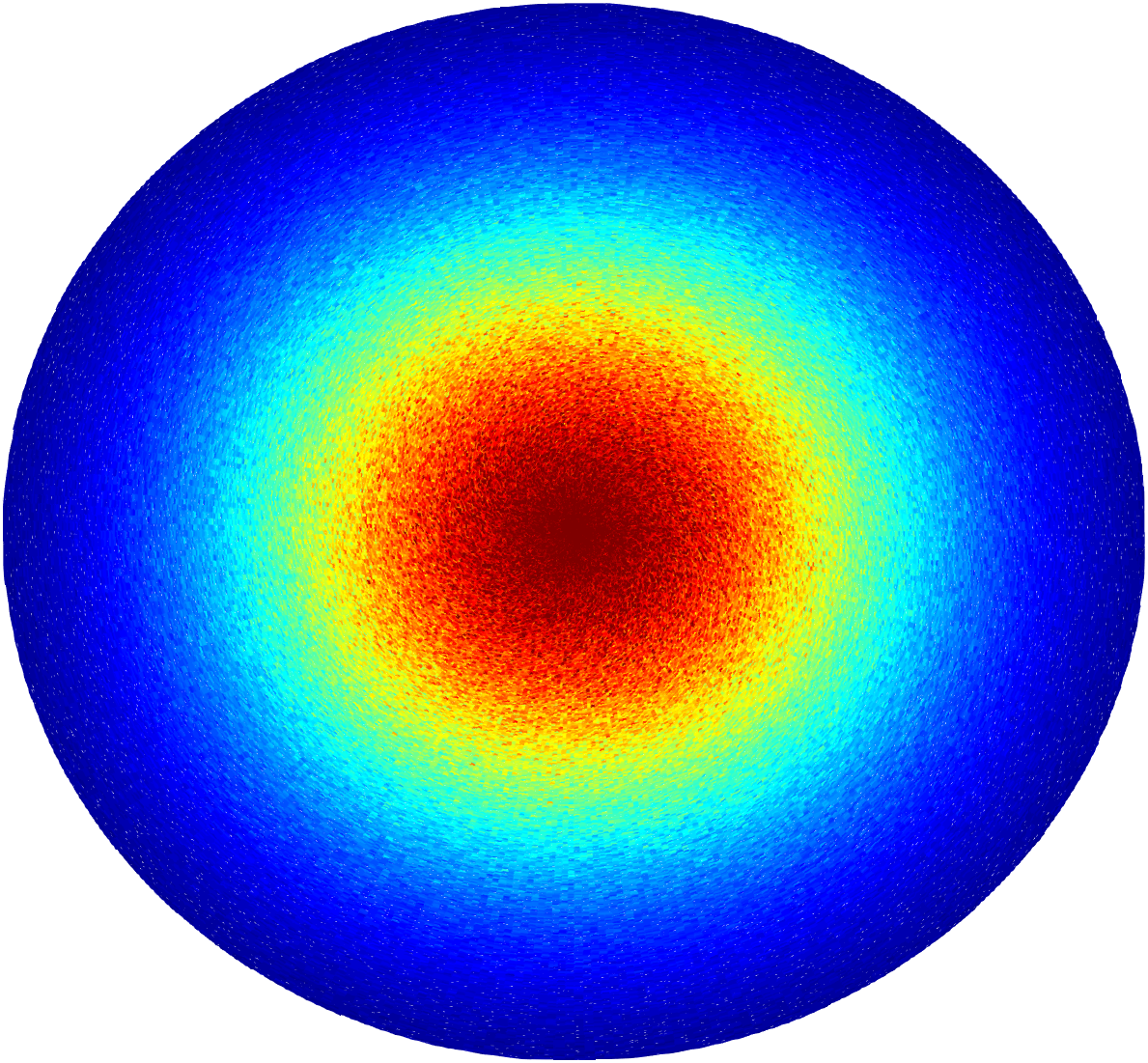}
	}
	\caption{(Color Online) (a) The spatial distribution of the light intensity of the beam cross-section without the scintillation effect. (b) The spatial distribution of the light intensity of the beam cross-section with the scintillation effect introduced. The intensity from weak to strong corresponds to the color from dark blue to dark red.}
	\label{ScintillationEffectSchematic}
\end{figure}

The scintillation effect is illustrated by the light intensity spatial distribution of the beam cross-section (see the schematic diagram in FIG. \ref{ScintillationEffectSchematic}). A great deal of turbulent vortices contained in the cross section independently scatters and diffracts the portion of the light impinging thereon, such that the intensity of light at each spatial point (irradiance) in the cross section varies randomly.

Over the years, many irradiance PDF models have been proposed to characterize the randomly fading irradiance signal, such as lognormal distribution, $K$ distribution, $I-K$ distribution, lognormal-Rician distribution, and gamma-gamma distribution \cite{Andrews2001}. These models are proposed for different turbulence strength regimes. The fluctuation strength is divided into weak and strong fluctuations corresponding to $\sigma_1^2<1$ and $\sigma_1^2>1$, respectively. Under weak fluctuations, the lognormal distribution is generally accepted, for Gaussian-beam wave it takes the form \cite{Andrews2001}
\begin{equation}
\begin{split}
\label{eq24}
p(I) = & \frac{1}{I\sigma_I( \textbf{r},L) \sqrt{2\pi}} \\
& \times\exp\left\lbrace -\frac{\left[\ln\left(\frac{I}{\langle I(\textbf{r},L)\rangle}\right) + \frac{1}{2}\sigma_I^2(\textbf{r},L) \right]^2 }{2\sigma_I^2(\textbf{r},L)} \right\rbrace, 
\end{split}
\end{equation}
where $\textbf{r}$ is a transverse vector, $\langle \cdot\rangle$ is an ensemble average, $\sigma_I^2(\textbf{r},L)$ is the scintillation index (for details see Appendix \ref{scintillationIndexWeak}), and $\langle I(\textbf{r},L)\rangle$ is the (normalized) mean irradiance (for details see Appendix \ref{MeanIrradiance}). Considering large-scale and small-scale effects, the (normalized) irradiance in strong fluctuation can be described by gamma-gamma distribution \cite{Andrews2001}
\begin{equation}
\label{eq25}
p(I) = \frac{2(\alpha\beta)^{(\alpha+\beta)/2}}{\Gamma(\alpha)\Gamma(\beta)}I^{(\alpha+\beta)/2-1}K_{\alpha-\beta}\left( 2\sqrt{\alpha\beta I}\right),
\end{equation}
where $\Gamma(\cdot)$ is the gamma function, $K_{\alpha-\beta}$ is the modified Bessel function of the second kind, $\alpha$ is the effective number of large scale cells of the scattering process, and $\beta$ is the effective number of small scale cells. Both $\alpha$ and $\beta$ are related to the scintillation index, and detailed in Appendix \ref{scintillationIndexStrong}.

Nevertheless, the distribution of irradiance only describes the intensity fluctuations at a certain spatial point. Hence, the irradiance should be integrated within the plane of the receiving aperture $\mathcal{B}$ :
\begin{equation}
P_{\rm rec} = \int_{\mathcal{B}}^{\,}I(\textbf{r},L){\rm d}\textbf{r}.
\end{equation}
Now, the transmittance can be written as
\begin{equation}
\label{eq46}
T_{\rm sci} = \frac{\int_{\mathcal{B}}^{\,}I(\textbf{r},L){\rm d}\textbf{r}}{\int_{\mathcal{A}}^{\,}I(\textbf{r},0){\rm d}\textbf{r}}
\end{equation}
where $\mathcal{A}$ is the plane of the transmitter aperture, and $I(\textbf{r},0)=\exp(-2r^2/W_0^2)$ is the (normalized) irradiance at the transmitter.

\begin{figure}[htbp]
	\centering
	\includegraphics[scale=0.55]{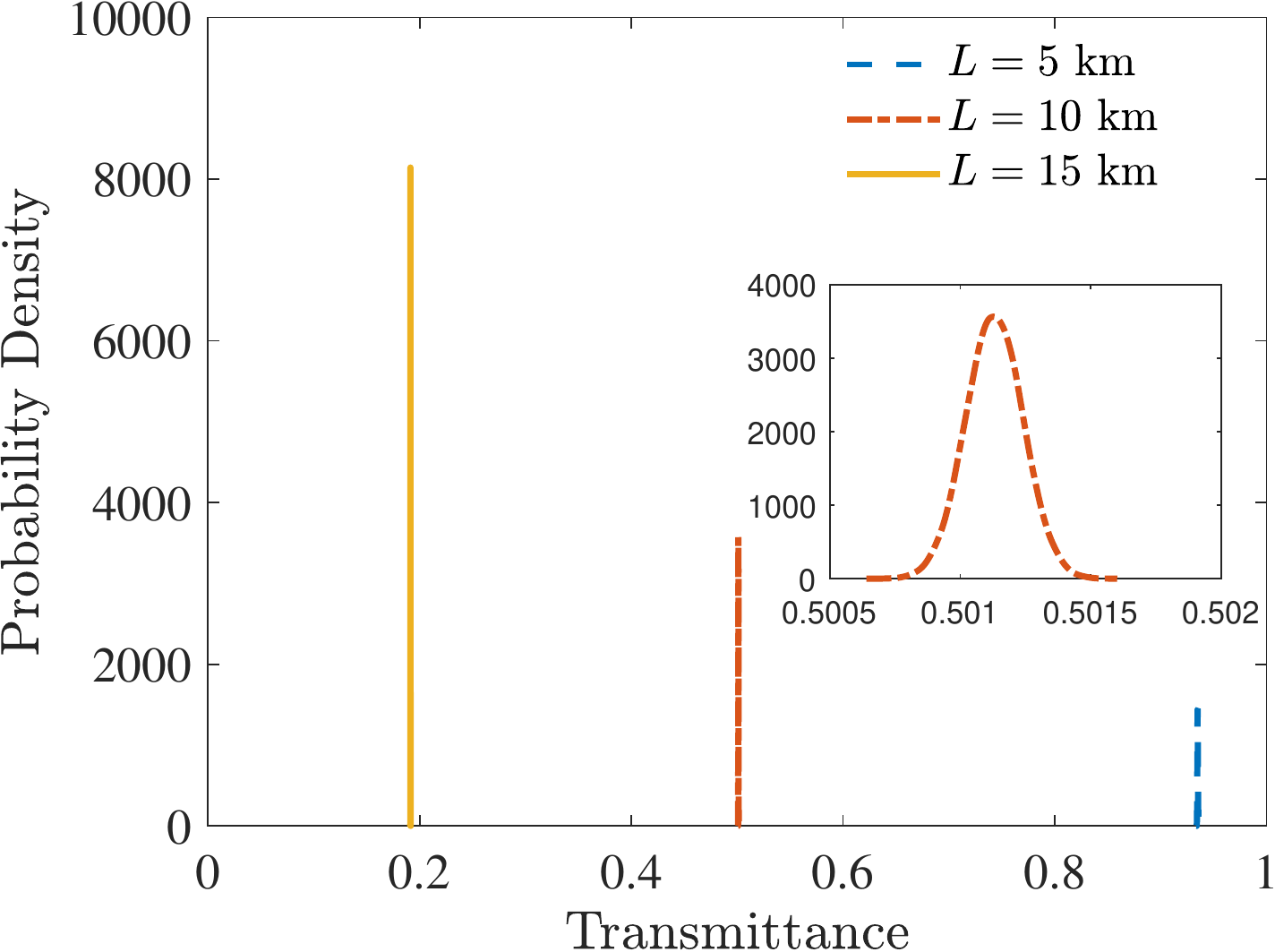}
	\caption{(Color Online) The probability density function of transmittance $T_{\rm sci}$ on the ground at distance 5, 10, and 15 km. The inset shows the PDF of transmittance at $L=10$ km. Further parameters: $a=110$ mm, $W_0=80$ mm.}
	\label{ScintillationPDF}
\end{figure}

Since $I(\textbf{r},L)$ is a random variable, it is quite difficult to calculate Eq.(\ref{eq46}) directly. Here, we still apply Monte Calro simulations to estimate $T_{\rm sci}$. For simplicity, only the scintillation and beam broadening reflected in $\left\langle I(\textbf{r},L)\right\rangle $ are considered, regardless of beam wandering and deformation. In this case, the scintillation-induced transmittance fluctuations is quite small, and the reduction in transmittance is primarily caused by the beam broadening, as shown in FIG. \ref{ScintillationPDF}. Since the scintillation-induced transmittance fluctuation is too small, the PDF in FIG. \ref{ScintillationPDF} looks just like a line. Therefore, the PDF of transmittance at $L=10$ km in the inset of FIG. \ref{ScintillationPDF} is presented as an example. The PDF shape of $L=5$ and 15 km are the same as the shape of $L=2$ km.

\subsection{Interruption Probability}
\label{secIP}
\begin{figure}[h]
	\centering
	\includegraphics[scale=0.55]{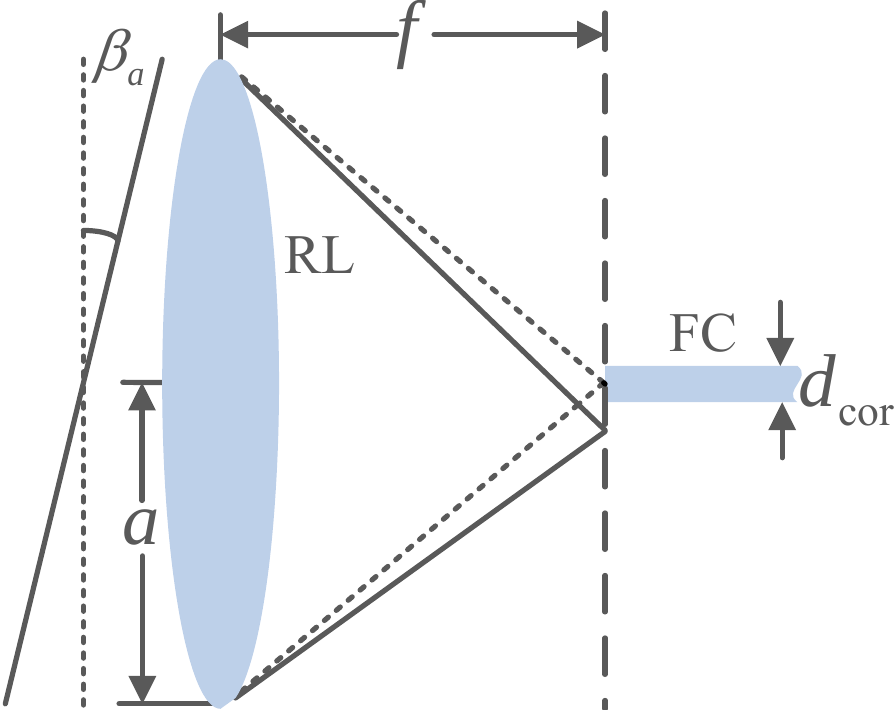}
	\caption{(Color Online) Communication interruption due to angle of arrival fluctuations. RL, receving lens; FC, fiber core. }
	\label{AngleOfArrival}
\end{figure}

Due to the high directivity of laser transmissions, there exists the possibility of communication interruption when there is a large angle-of-arrival fluctuation. The direct reflection of angle-of-arrival fluctuations on the receiving aperture plane is image jitter on a focal plane. When the focus is not within the receiving fiber core, communication is interrupted at this time (see FIG. \ref{AngleOfArrival}).

\begin{figure}[h]
\centering
\includegraphics[scale=0.55]{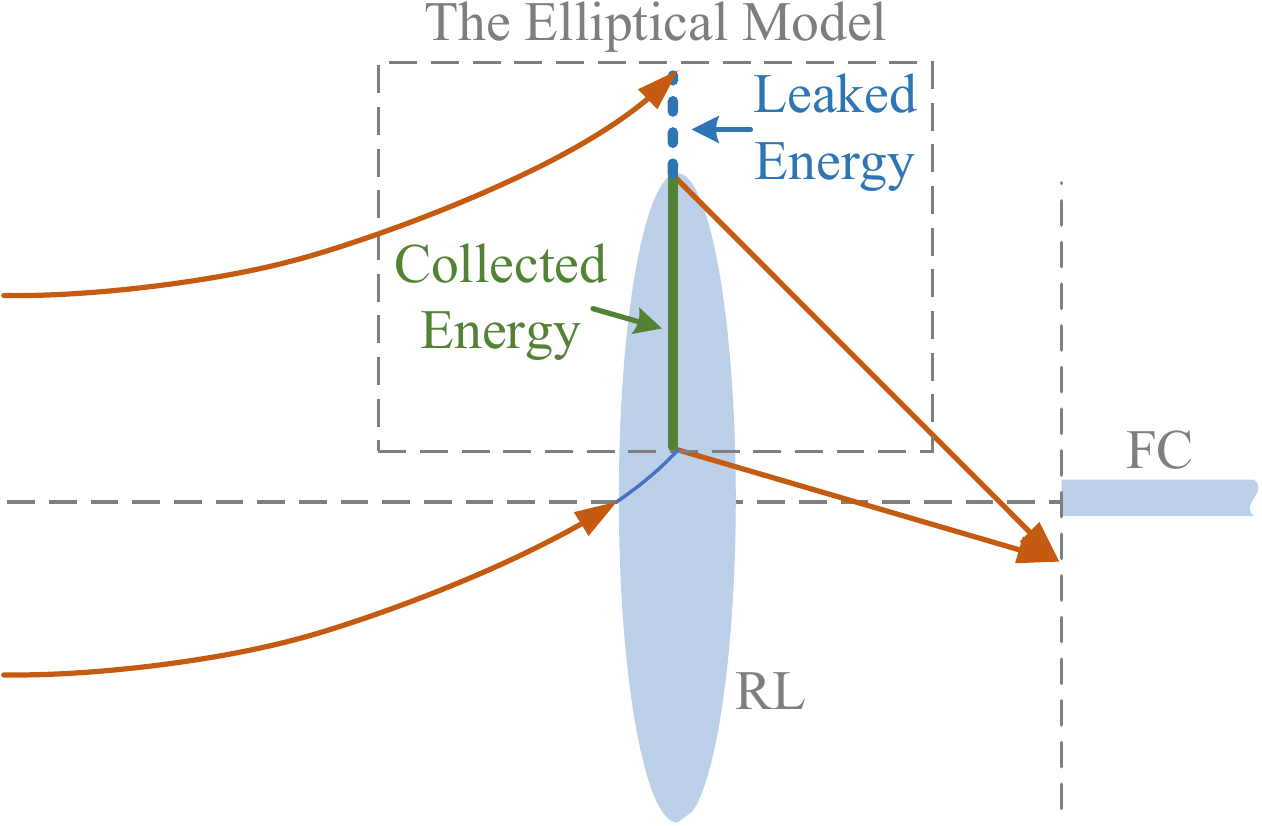}
\caption{(Color Online) Relation between the elliptical beam model and communication interruption.}
\label{EB-AOA}
\end{figure}

The interruption phenomenon is closely related to beam wandering comprised in the elliptical model. However, the major concern of the elliptical beam model is the total energy collected by RL, i.e., truncation of beam spot. It is still possible that part of the beam spot is within RL while the focus displaced out of FC, as illustrated in Fig. \ref{EB-AOA}. At this point, if only the elliptical beam model is involved, the signal transmission will still be considered as successful, but in fact not. Thus, the consideration of communication interruption is necessary. Due to the close relationship between beam wandering and interruption, the statistics of them should be the same, that is, $\beta_a$ is Gaussian distributed.

Assuming the mean value of arriving angle $\langle\beta_a\rangle = 0$ and $\beta_a$ is small enough so that $\sin\beta_a\cong\beta_a$, the variance of $\beta_a$ can be written as
\begin{equation}
\label{eq47}
\langle \beta_a^2 \rangle = \frac{\left\langle \Delta x_0^2\right\rangle }{L^2},
\end{equation}
where $\left\langle \Delta x_0^2\right\rangle$ is further expressed in TABLE \ref{tab1}. The rms image displacement is then given as
\begin{equation}
\label{eq50}
L_{\rm dis} = f\sqrt{\langle \beta_a^2 \rangle}
\end{equation}
where $f$ is the focal length of the collecting lens.

\begin{figure}[h]
	\centering
	\includegraphics[scale=0.55]{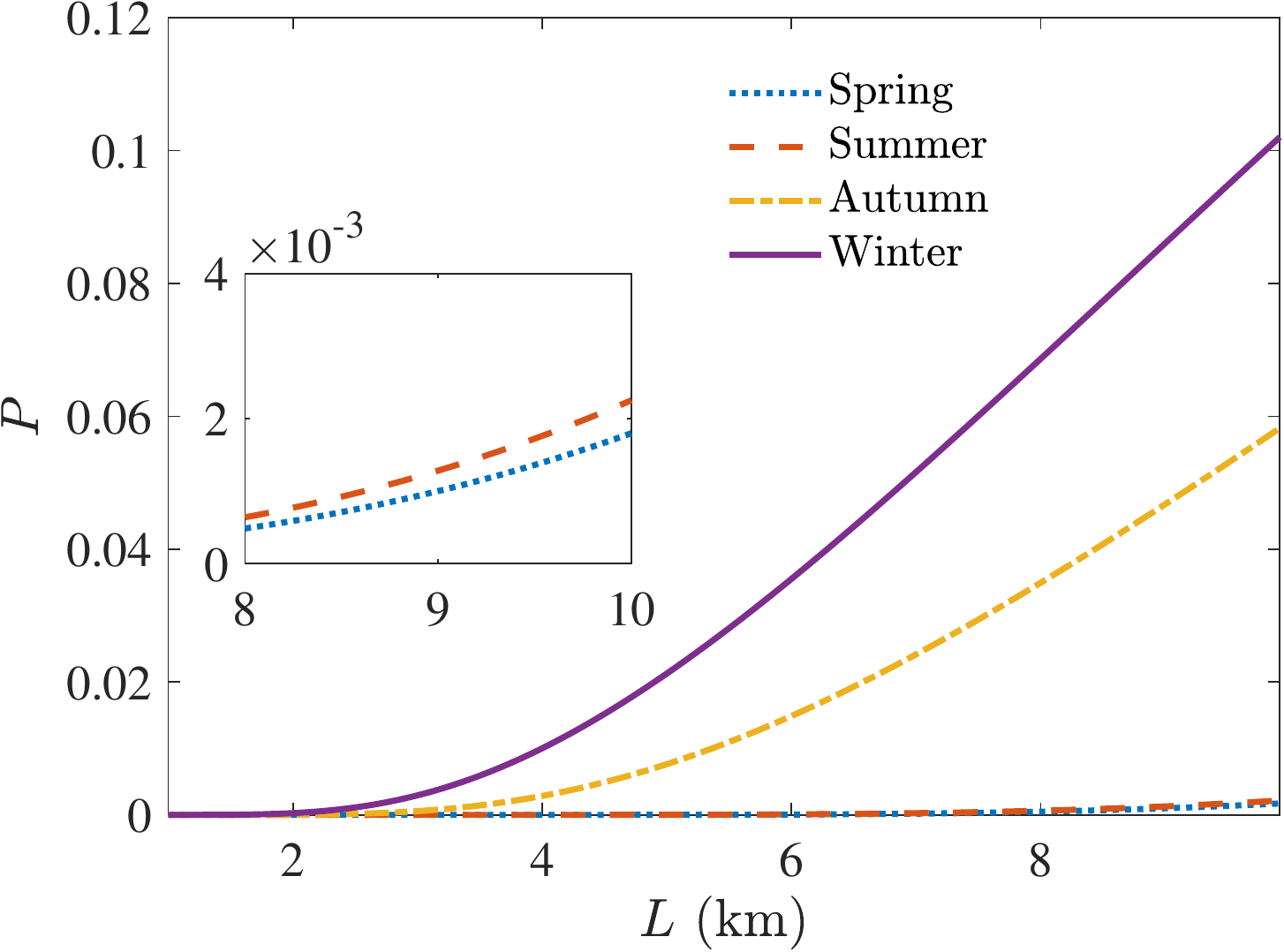}
	\caption{(Color Online) The communication interruption probability of four seasons varies with distance. The inset shows probability of spring and summer from 8 to 10 km. Further parameters: fiber core diameter 9 $\mu$m, $f=220$ mm, $a=110$ mm, and $W_0=80$ mm.}
	\label{IP}
\end{figure}

Since $\beta_a$ is normally distributed, the focus displacement is also normally distributed. Thus, the communication interruption probability can be expressed as
\begin{equation}
P = 1 - \int_{-d_{\rm cor}/2}^{d_{\rm cor}/2} \frac{1}{\sqrt{2\pi\langle \beta_a^2 \rangle}f} \exp\left( \frac{-l^2}{2f^2\langle \beta_a^2 \rangle} \right) {\rm d}l,
\end{equation}
where $d_{\rm cor}$ is the diameter of the fiber core in meters. A typical single-mode optical fiber has a core diameter from 8.3 to 10.5 $\mu$m. Here, as an illustration, we assume that the core diameter is 9 $\mu$m, $f=220$ mm, $a=110$ mm, and $W_0=80$ mm. The interruption probability is shown in Fig. \ref{IP}.

\subsection{Excess Noise}
\label{secnoise}
\begin{figure}
	\centering
	\includegraphics[scale=1]{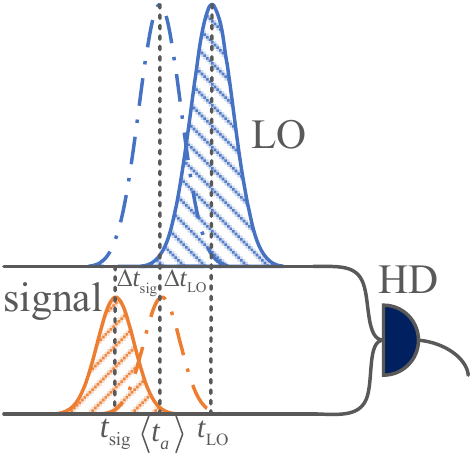}
	\caption{(Color Online) The arrival time fluctuation causes the LO and signal not to be aligned in time domain when interfering.}
	\label{ArrivalTime}
\end{figure}

In this subsection we will focus on pulse arrival time fluctuations observed by a fixed observer (see FIG. \ref{ArrivalTime}). This effect may bring extra excess noise by causing phase fluctuations.

The pulse arrival time between the LO and signal $\Delta t$ is now a random variable. To clarify $\Delta t$, we first investigate the arrival time of a single pulse $t_a$. The mean value and variance of $t_a$ are given by \cite{Young1998}
\begin{equation}
	\langle t_a \rangle =  \frac{\left\langle M^{(1)} \right\rangle }{\left\langle M^{(0)} \right\rangle}, \;
	\langle t_a^2 \rangle =  \frac{\left\langle M^{(2)} \right\rangle }{\left\langle M^{(0)} \right\rangle}, \;
	\sigma_{\rm ta}^2 =  \langle t_a^2 \rangle - \langle t_a \rangle^2,
\end{equation}
where
\begin{equation}
\left\langle M^{(n)} \right\rangle = \int_{-\infty}^{\infty}t^n\left\langle v_0(\textbf{r},L;t)v_0^\ast(\textbf{r},L;t) \right\rangle {\rm d} t
\end{equation}
is the $n$-th moment with the complex envelope $v_0(\textbf{r},L;t)$. Under weak turbulence, near-field and far-field approximations, the mean value and on-axis variance of arrival time is given by \cite{Andrews2005}
\begin{equation}
\langle t_a \rangle = \frac{L}{c}, \quad \sigma_{\rm ta}^2 = \frac{T_1^2}{4},
\end{equation}
where $T_1$ is given in Eq.(\ref{T1}). The on-axis variance of strong turbulence is deduced by \cite{Chen2012}.

Now we define the random variable $\Delta t$ as
\begin{equation}
\Delta t = t_{\rm LO} - t_{\rm sig},
\end{equation}
where $t_{\rm LO} = \langle t_a \rangle + \Delta t_{\rm LO}$ and $t_{\rm sig} = \langle t_a \rangle + \Delta t_{\rm sig}$ are shown in FIG. \ref{ArrivalTime}. This directly leads to
\begin{equation}
\Delta t = \Delta t_{\rm LO} - \Delta t_{\rm sig},
\end{equation}
where $\Delta t_{\rm LO}$ and $\Delta t_{\rm sig}$ are random variables with mean value zero and variance $\sigma_{\rm ta}^2$. Thus,
\begin{equation}
\label{Deltat}
\left\langle \Delta t \right\rangle = 0, \quad \sigma_{\Delta t}^2 = 2(1-\rho_{\rm ta})\sigma_{\rm ta}^2,
\end{equation}
where $\left\langle \Delta t \right\rangle$ and $\sigma_{\Delta t}^2$ are the mean value and variance of $\Delta t$, respectively. Here $\rho_{\rm ta}$ is the correlation coefficient between $\Delta t_{\rm LO}$ and $\Delta t_{\rm sig}$.

Now we can deduce the variance of phase fluctuation with Eq.(\ref{Deltat})
\begin{equation}
\sigma_{\theta}^2 = \omega^2 \sigma_{\Delta t}^2,
\end{equation}
where $\omega$ is the angular frequency of light mentioned in Eq.(\ref{freespaceFar}). When the phase fluctuation is small enough, the excess noise can be expressed as \cite{Qi2015}
\begin{equation}
\varepsilon_\theta = V_A \sigma_{\theta}^2,
\end{equation}
where $V_A = V - 1$ is the modulation variance of Alice.

\begin{figure}
	\centering
	\includegraphics[scale=0.52]{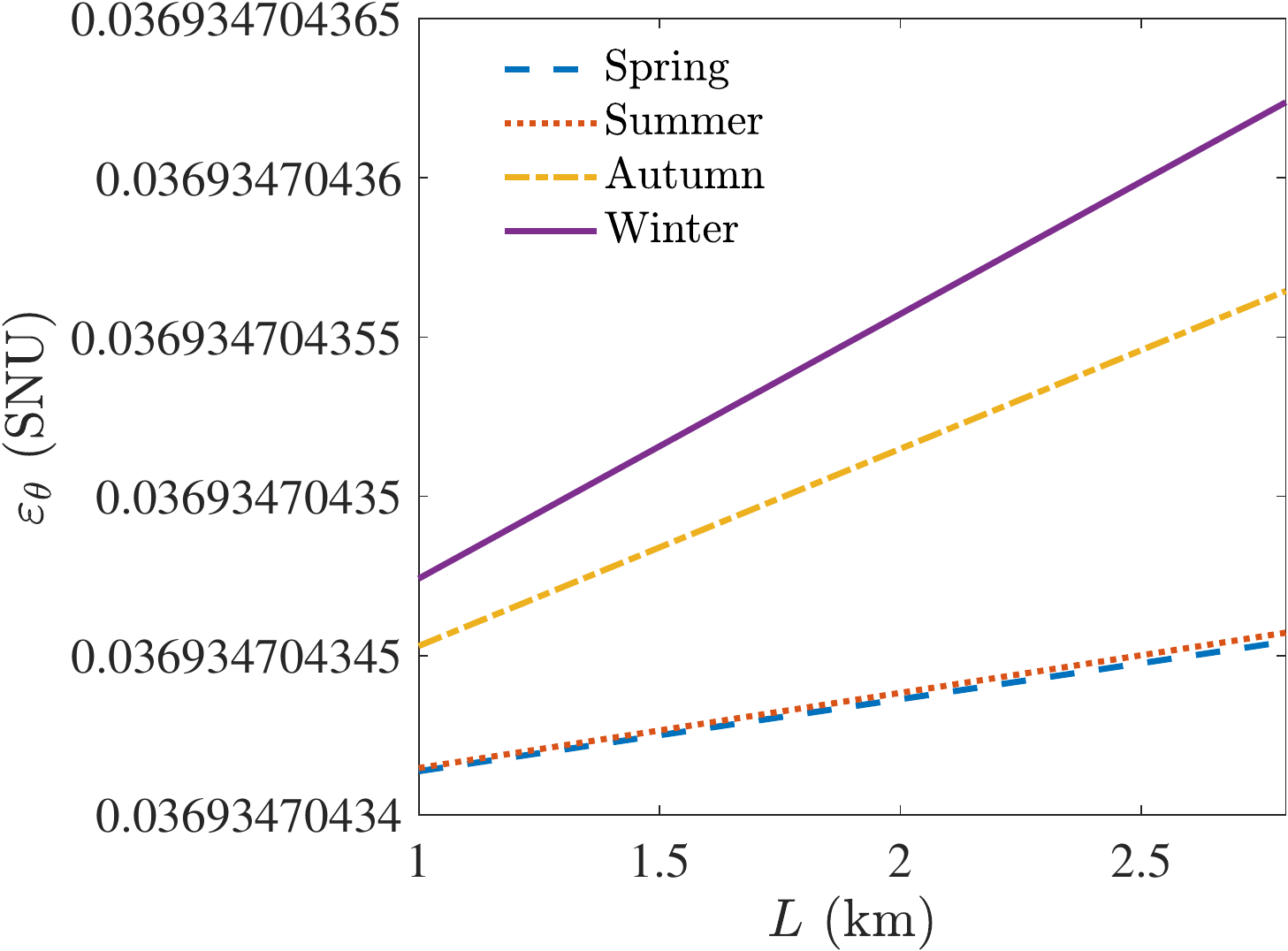}
	\caption{(Color Online) The excess noise caused by phase fluctuations.}
	\label{ExcessNoise}
\end{figure}

FIG. \ref{ExcessNoise} shows the excess noise varies with distance $L$. Here we consider that $f_{\rm PRF} = 100$ MHz, $R_{dut}=10\%$, $\rho_{\rm ta} = 1-10^{-13}$, and $V_A = 2$ in shot noise unit (SNU). With such a high correlation coefficient and weak turbulence condition, phase excess noise can eventually reach an acceptable level, which is hard to achieve in practice. The phase compensation method for fiber CVQKD \cite{Huang2016} can compensate small phase fluctuations, but it is not applicable to atmosphere CVQKD whose phase fluctuations is very large. Therefore, we hope that an effective phase compensation method for atmospheric CVQKD will be proposed in the future. It is also worth noting that the turbulence-induced phase excess noise may be more readily to be decreased, if a non-pilot aided "local" LO scheme, which does not require reference pulses or pilot, e.g. the experiment \cite{Günthner2017}, is successfully applied in GMCS CVQKD. This is because only the arrival time of the signal is fluctuant, whereas the LO is not transmitted through the channel. Unfortunately, in this scheme, part of signal would be split off to lock phase. This operation would bring extra noise and decrease total detection efficiency in a complete experiment. Therefore, it would be a trade-off between the simultaneous transmission and "local" LO scheme.

\section{performance analysis}
\label{sec4}
In this section, we will combine the results of section \ref{sec2} and \ref{sec3} to analyze the achievable final key rate. 

The Monte Carlo method is applied to estimate the secret key rate in Eq.(\ref{eq14}). Since the excess noise caused by phase fluctuations can not yet be accurately compensated, it is quite difficult to estimate the practical excess noise after using different effective compensation methods. Thus, here we do not consider the phase excess noise that changes with atmospheric conditions for the time being, but we still examine the achievable key rate under different fixed excess noise level, namely $\varepsilon=0.01$ and $\varepsilon=0.03$ in SNU. And as explained at the end of section \ref{secTPB}, the temporal pulse broadening will not be considered in this section. The extinction coefficients used are listed in TABLE \ref{TableExt} and all other parameters needed in performance analysis are presented in TABLE \ref{tab3}.

\begin{table}[h]
	\centering
	\caption{The parameters setting for performance analysis.}
	\label{tab3}
	\begin{tabular}{ccc}
		\toprule[0.75pt]
		Parameters & Values & Description \\
		\midrule[0.5pt]
		$a$ & 110 mm & Receiving lens radius \\
		$W_0$ & 80 mm & Transmitting lens radius \\
		$f$ & 220 mm & Focal length of receiving lens \\
		$d_{\rm cor}$ & 9 $\mu$m & Fiber core diameter \\
		$l_0$ & 4 mm & Inner scale of atmosphere \\
		$L_0$ & 0.4 m & Outer scale of atmosphere \\
		$\lambda$ & 1550 nm & Laser wavelength \\
		$V_A$ & 2 SNU & Modulation variance\\
		$\upsilon_{el}$ & 0.01 SNU & Electronic noise\\
		$\varepsilon$ & 0.01; 0.03 SNU & Excess noise\\
		$\eta$ & $60\%$ & Detection efficiency\\
		$\beta$ & $90\%$ & Reconciliation efficiency\\
		\bottomrule[0.75pt]
	\end{tabular}
\end{table}

\begin{figure}[h]
	\centering
	\subfigure[]{
		\includegraphics[scale=0.55]{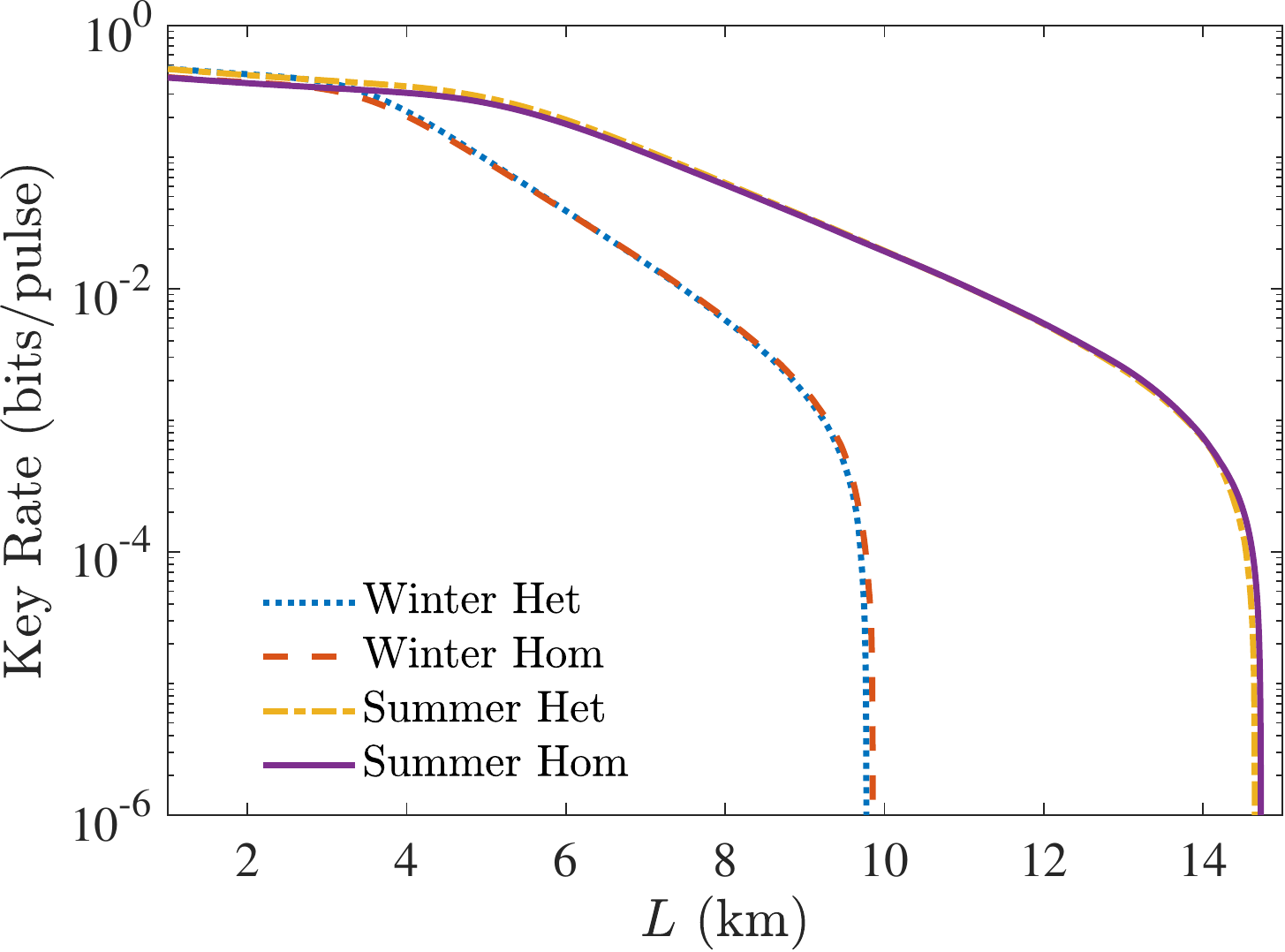}
		\label{epsilon001}
	}
	\subfigure[]{
		\includegraphics[scale=0.55]{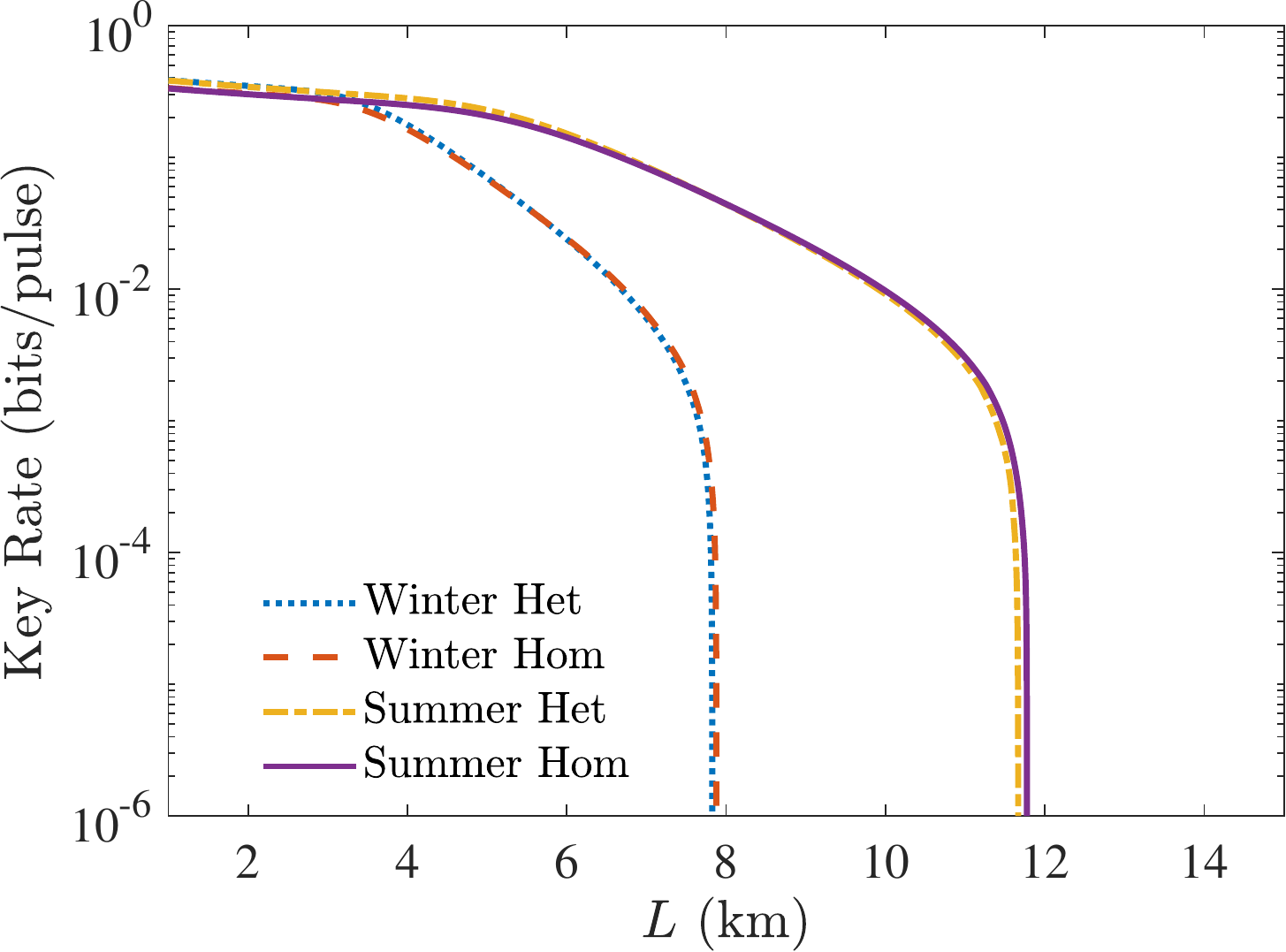}
		\label{epsilon003}
	}
	\caption{(Color Online) The secret key rate as a function of distance for homodyne and heterodyne in summer and winter. (a) $\varepsilon=0.01$. (b) $\varepsilon=0.03$.}
	\label{performance}
\end{figure}

The secret key rate with $\varepsilon=0.01$ SNU is demonstrated in FIG. \ref{epsilon001}. As we can see, the transmission distance of system using homodyne detection is longer than heterodyne detection, but not much. Now we increase the excess noise to $\varepsilon=0.03$ SNU, as depicted in FIG. \ref{epsilon003}. Compared with FIG. \ref{epsilon001}, the achievable transmission distance of is obviously shorter.

The performance analysis conducted in this section indicates several key points. First, homodyne detection should be applied to practical systems if farther transmission distance is demanded, otherwise, heterodyne provides higher achievable key rate at short distance. Here, especially, the heterodyne case deserves attentions, since only the no-switching protocol \cite{Weedbrook2004} had been proven to against general attacks in a realistic finite-size regime \cite{Leverrier2013, Leverrier2017}. Second, we found that the transmittance fluctuations are destructive to the key rate. Accordingly, in practical experiments, the main effort should be devoted to inhibiting the effects of beam wandering, broadening and deformation. Third, since the impact of excess noise is quite significant, and the phase excess noise would be much more than 0.03, effective methods of controlling phase excess noise are needed to increase the secret key rate. It is noteworthy that our performance analysis was conducted on the assumption that there is no enhancement technique, such as adaptive optics \cite{Tyson2010,Günthner2017} and post selection \cite{Usenko2012}. Adaptive optics are shown to be cable of relieving signal fading \cite{Tyson2003}, correcting wavefront and improving fiber coupling efficiency \cite{Wilks2002}.  The incoming result of post selection is, actually, increasing $\langle T\rangle$ by discarding data when instantaneous transmittance is too low. These methods or techniques can directly or indirectly weaken atmospheric effects. Therefore, the performance of practical systems may be better than the results shown in this paper. We expect that phase compensation techniques for atmospheric CVQKD can also be proposed so that the phase excess noise can be reduced to a negligible level.

\section{Conclusion}
\label{sec5}
We analyzed atmospheric effects on the horizontal CVQKD links on the Earth's surface, thus establishing a transmission model, which can help the performance assessment of practical CVQKD systems. The newly derived key rate formulas for fading channels with detection efficiency and noise considered shows that there are three main parameters that affect the final key rate: the transmittance, interruption probability and excess noise. The transmittance change caused by temporal pulse broadening under weak turbulence regime is found to be negligible. Transmittance fluctuations caused by beam wandering, broadening, deformation, and scintillation make the final key rate deteriorated rapidly. Angle-of-arrival fluctuations may cause communication interruptions which leads to a more obvious decline in the key rate over long-distance transmission. The phase excess noise caused by pulse arrival time fluctuation is found to be quite large. We found, in fading channels, systems employing homodyne detection can transmit far more distances than heterodyne detection, while employment of heterodyne detection at short-range transmission has a higher key rate than homodyne detection.

\section*{Acknowledgements}
This work was supported by the National Natural Science Foundation of China (Grants No. 61332019, 61671287, 61631014), and the National key research and development program (Grant No. 2016YFA0302600).

\appendix
\section{the Symplectic Eigenvalues of the Holevo Quantity}
\label{appendixA}
The symplectic eigenvalues $\lambda_{1,2}$ can be calculated for both homodyne and heterodyne detection by
\begin{equation}
\lambda_{1,2}^2 = \frac{1}{2}\left[A\pm\sqrt{A^2-4B}\right],
\end{equation}
with
\begin{equation}
\begin{split}
A = & V^2\left(1-2\langle\sqrt{T}\rangle^2\right) + 2\langle\sqrt{T}\rangle^2 \\ 
& + \langle T\rangle^2\left(V+1/\langle T\rangle-1+\varepsilon\right)^2, \\
B = & \bigg[V^2{\rm var}\left(\sqrt{T}\right) + \langle\sqrt{T}\rangle^2 \\
& + \langle T\rangle V(1/\langle T\rangle-1+\varepsilon)\bigg]^2,
\end{split}
\end{equation}
where ${\rm var}\left(\sqrt{T} \right) = \langle T\rangle - \langle\sqrt{T}\rangle^2$ is the variance of $\sqrt{T}$. Then, $\lambda_{3,4,5}$ are the symplectic eigenvalues of covariance matrix $\gamma_{\rm AFG}^{m_{\rm B}}$, which can be expressed as
\begin{equation}
\label{eq9}
\gamma_{\rm AFG}^{m_{\rm B}} = \gamma_{\rm AFG} - \gamma^{\rm h},
\end{equation}
where $\gamma^{\rm h} = \sigma_{\rm AFGB_2}^{T}H\sigma_{\rm AFGB_2}$. For homodyne, $H^{\rm hom}=(X\gamma_{\rm B_3}X)^{\rm MP}$, where $X=diga(1,0)$ and MP represents Moore-Penrose pseudo-inverse, for heterodyne, $H^{\rm het}=(\gamma_{\rm B_3}+\mathbb{I})^{-1}$. The covariance matrix of four modes
\begin{equation}
\gamma_{\rm AFGB_3}=
\begin{pmatrix}
\gamma_{\rm AFG} & \sigma_{\rm AFGB_2}^{T} \\\
\sigma_{\rm AFGB_2} & \gamma_{\rm B_3} 
\end{pmatrix}
\end{equation} 
comprises all the elements. To simplify the results of Eq.(\ref{eq9}), we define elements of Eq.(\ref{eq4}) as
\begin{equation}
\begin{split}
a = & V, \\
b = & \langle\sqrt{T}\rangle\sqrt{V^{2}-1}, \\
c = & \langle T\rangle(V+1/\langle T\rangle-1+\varepsilon),
\end{split}
\end{equation}
then we can deduce
\begin{equation}
\label{eq11}
\gamma_{\rm AFG} = 
\begin{pmatrix}
a\mathbb{I} & b\sqrt{1-\eta}\sigma_{z} & \textbf{0} \\
b\sqrt{1-\eta}\sigma_{z} & [(\nu-c)\eta+c]\mathbb{I} & \sqrt{\eta(\nu^2-1)}\sigma_{z} \\
\textbf{0} & \sqrt{\eta(\nu^2-1)}\sigma_{z} & \nu\mathbb{I}
\end{pmatrix}
\end{equation}
for both homodyne and heterodyne detection, and
\begin{equation}
\label{eq12}
\gamma^{\rm hom} = \frac{1}{(c-\nu)\eta+\nu}
\begin{pmatrix}
\gamma_{1-1}^{\rm hom} & \gamma_{2-1}^{\rm hom} & \gamma_{3-1}^{\rm hom} \\
\gamma_{2-1}^{\rm hom} & \gamma_{2-2}^{\rm hom} & \gamma_{3-2}^{\rm hom} \\
\gamma_{3-1}^{\rm hom} & \gamma_{3-2}^{\rm hom} & \gamma_{3-3}^{\rm hom}
\end{pmatrix},
\end{equation}
with
\begin{equation}
\label{eq13}
\begin{split}
&\gamma_{1-1}^{\rm hom} = b^2\eta X, \\
&\gamma_{2-2}^{\rm hom} = (c-\nu)^2\eta(1-\eta)X, \\
&\gamma_{3-3}^{\rm hom} = (1-\eta)(\nu^2-1)X, \\
&\gamma_{2-1}^{\rm hom} = b\eta(c-\nu)\sqrt{1-\eta}X, \\
&\gamma_{3-1}^{\rm hom} = -b\sqrt{\eta(1-\eta)(\nu^2-1)}X, \\
&\gamma_{3-2}^{\rm hom} = -(c-\nu)(1-\eta)\sqrt{\eta(\nu^2-1)}X,
\end{split}
\end{equation}
for homodyne case, where $\nu=1+\upsilon_{el}/(1-\eta)$ and $X = diag(1,0)$, while for heterodyne case
\begin{equation}
\label{Chihet}
\gamma^{\rm het} = \frac{1}{(c-\nu)\eta+\nu+1}
\begin{pmatrix}
\gamma_{1-1}^{\rm het} & \gamma_{2-1}^{\rm het} & \gamma_{3-1}^{\rm het} \\
\gamma_{2-1}^{\rm het} & \gamma_{2-2}^{\rm het} & \gamma_{3-2}^{\rm het} \\
\gamma_{3-1}^{\rm het} & \gamma_{3-2}^{\rm het} & \gamma_{3-3}^{\rm het}
\end{pmatrix},
\end{equation}
with
\begin{equation}
\label{gammas_het}
\begin{split}
&\gamma_{1-1}^{\rm het} = b^2\eta \mathbb{I}, \\
&\gamma_{2-2}^{\rm het} = (c-\nu)^2\eta(1-\eta) \mathbb{I}, \\
&\gamma_{3-3}^{\rm het} = (1-\eta)(\nu^2-1) \mathbb{I}, \\
&\gamma_{2-1}^{\rm het} = b\eta(c-\nu)\sqrt{1-\eta} \sigma_{z}, \\
&\gamma_{3-1}^{\rm het} = -b\sqrt{\eta(1-\eta)(\nu^2-1)} \mathbb{I}, \\
&\gamma_{3-2}^{\rm het} = -(c-\nu)(1-\eta)\sqrt{\eta(\nu^2-1)} \sigma_{z},
\end{split}
\end{equation}
where $\nu=1+2\upsilon_{el}/(1-\eta)$. Substituting Eq.(\ref{eq11}), (\ref{eq12}) and (\ref{Chihet}) into Eq.(\ref{eq9}) yields the final result of $\gamma_{\rm AFG}^{m_{\rm B}}$. Then, we can calculate $\lambda_{3,4,5}$ through $\gamma_{\rm AFG}^{m_{\rm B}}$. The symplectic eigenvalues $\lambda_{3,4}$ can take the same form as
\begin{equation}
\lambda_{3,4}^2 = \frac{1}{2}\left[ C\pm \sqrt{C^2-4D}\right]
\end{equation}
for both homodyne and heterodyne case, while $\lambda_{5}$ is found to be 1. Specifically, $C$ and $D$ for homodyne and heterodyne case can be expressed as
\begin{equation}
\begin{split}
& C_{\rm hom} = \frac{A\chi_{\rm hom}+a\sqrt{B}+c}{c+\chi_{\rm hom}}, \\
& D_{\rm hom} = \sqrt{B}\frac{a+\sqrt{B}\chi_{\rm hom}}{c+\chi_{\rm hom}},
\end{split}
\end{equation}
and
\begin{equation}
\begin{split}
& C_{\rm het} = \frac{A\chi_{\rm het}^2+2\chi_{\rm het}(a\sqrt{B}+c)+B+2b^2+1}{(c+\chi_{\rm het})^2}, \\
& D_{\rm het} = \left(\frac{a+\sqrt{B}\chi_{\rm het}}{c+\chi_{\rm het}}\right)^2,
\end{split}
\end{equation}
respectively, where $\chi_{\rm hom}=(1-\eta+\upsilon_{el})/\eta$ and $\chi_{\rm het}=(2-\eta+2\upsilon_{el})/\eta$ stand for the detection-added noise (SNU) of homodyne and heterodyne detection, respectively.

\section{the Estimation of $\left\langle \sqrt{T_{\rm bro}} \right\rangle$}
\label{Tbro}
In this appendix we estimate the value of $\left\langle \sqrt{T_{\rm bro}} \right\rangle$ when $\left\langle T_{\rm bro} \right\rangle \cong 1$.

\begin{figure}
	\centering
	\includegraphics[scale=0.55]{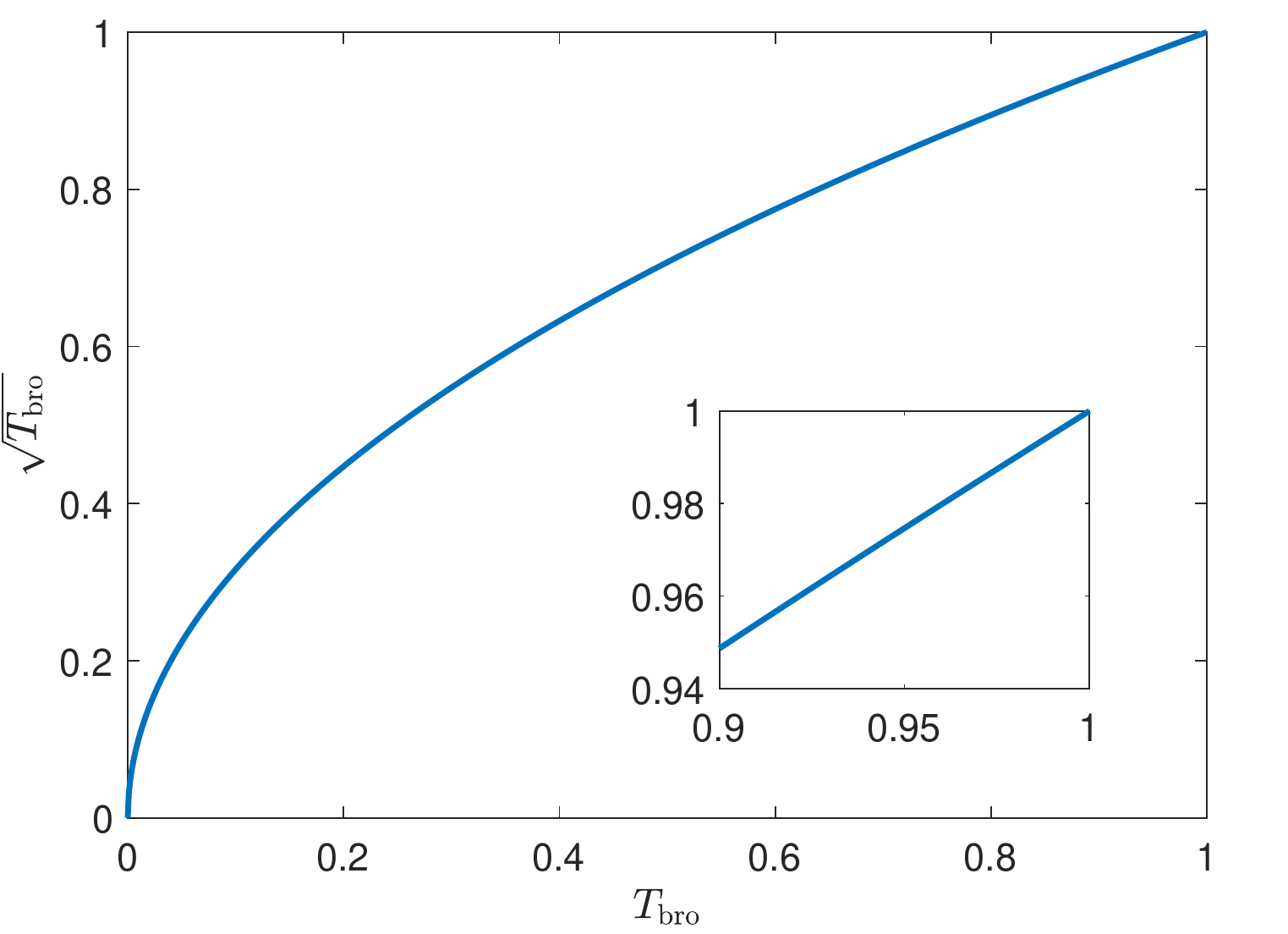}
	\caption{The relationship between $\sqrt{T_{\rm bro}}$ and $T_{\rm bro}$. The inset shows the details of the relationship from $T_{\rm bro}=0.9$ to 1.}
	\label{BroadeningEstimation}
\end{figure}

Although both $\sqrt{T_{\rm bro}}$ and $T_{\rm bro}$ are random variables, the relationship between them can be determined, as shown in FIG. \ref{BroadeningEstimation}.
Since $\left\langle T_{\rm bro} \right\rangle \cong 1$ and $T_{\rm bro}\leq1$, only the area around $T_{\rm bro}\cong1$ needs to be considered. We can see from the inset in FIG. \ref{BroadeningEstimation} that when $T_{\rm bro} \cong1$, the relationship between $\sqrt{T_{\rm bro}}$ and $T_{\rm bro}$ is approximately linear, namely,
\begin{equation}
\label{eqB1}
\sqrt{T_{\rm bro}} \cong k_{\rm bro} T_{\rm bro} + c_{\rm bro},
\end{equation}
where $k_{\rm bro}$ is the slope and $c_{\rm bro}$ is a constant, and there exists the relationship $k_{\rm bro}+c_{\rm bro}=1$ which can be obtained by setting $T_{\rm bro}=1$ in Eq. (\ref{eqB1}). This immediately leads to
\begin{equation}
\left\langle \sqrt{T_{\rm bro}} \right\rangle \cong k_{\rm bro} \left\langle T_{\rm bro} \right\rangle + c_{\rm bro}.
\end{equation}
With $\left\langle T_{\rm bro} \right\rangle \cong 1$, we can come to the conclusion that $\left\langle \sqrt{T_{\rm bro}} \right\rangle \cong 1$.

\section{the Parameters of $T_{\rm ell}$}
\label{ELLparameters}
$W_{\rm eff}(\zeta)$ is the effective squared spot radius expressed as
\begin{equation}
\begin{split}
W_{\rm eff}(\zeta) = & 2a\Bigg[ \mathcal{W}\bigg( e^{(a^2/W_1^2)[1+2\cos^2\zeta]} \\
& \times\frac{4a^2}{W_1W_2}e^{(a^2/W_2^2)[1+2\sin^2\zeta]} \bigg) \Bigg]^{-\frac{1}{2}}
\end{split}
\end{equation}
with the Lambert $W$ function $\mathcal{W}(\cdot)$ \cite{Corless1996}. $T_{\rm ell,0}$ is the transmittance of the centered beam ($r_0=0$) given as
\begin{equation}
\begin{split}
T_{\rm ell,0} = & 1 - I_0\left( a^2\left[ \frac{1}{W_1^2}-\frac{1}{W_2^2} \right] \right) e^{-a^2\left(\frac{1}{W_1^2}+\frac{1}{W_2^2}\right)} \\
& -2\left\lbrace  1-\exp\left[ -\frac{a^2}{2}\left( \frac{1}{W_1}-\frac{1}{W_2} \right)^2 \right] \right\rbrace \\
& \times\exp\left\lbrace -\left[ \frac{\frac{(W_1+W_2)^2}{W_1^2-W_2^2}}{R\left(\frac{1}{W_1}-\frac{1}{W_2}\right)} \right]^{Q\left(\frac{1}{W_1}-\frac{1}{W_2}\right)} \right\rbrace ,
\end{split}
\end{equation}
with the modified Bessel function of $i$th order $I_i(\xi)$, the scale function
\begin{equation}
R(\xi) = \left[ \ln\left(2\frac{1-\exp(-a^2\xi^2/2)}{1-\exp(-a^2\xi^2)I_0(a^2\xi^2)} \right) \right]^{-\frac{1}{Q(\xi)}}
\end{equation}
and the shape function
\begin{equation}
\label{eqQ}
\begin{split}
Q(\xi) = & 2a^2\xi^2\frac{\exp(-a^2\xi^2)I_1(a^2\xi^2)}{1-\exp(-a^2\xi^2)I_0(a^2\xi^2)} \\
& \times\left[ \ln\left( 2\frac{1-\exp(-a^2\xi^2/2)}{1-\exp(-a^2\xi^2)I_0(a^2\xi^2)} \right) \right]^{-1}.
\end{split}
\end{equation}

\section{Mean Values and Covariance Matrix Elements of $\textbf{v}$}
\label{TABLE}
The mean values and covariance matrix elements of $\textbf{v}$ is shown in TABLE \ref{tab1}, where $\Omega=kW_0^2/2L$ is the Fresnel parameter, and $\gamma=(1+\Omega^2)/\Omega^2$. These mean values and elements are given for horizontal links.
\begin{table}[htbp]
	\centering
	\caption{Mean values and elements of the covariance matrix of $\textbf{v}$ for horizontal link.}
	\label{tab1}
	\begin{tabular}{cc}
		\toprule[0.75pt]
		& Weak Turbulence \\
		\midrule[0.5pt]
		$\langle \Theta_{1,2} \rangle$ & $\ln\left[ \frac{\left(1+2.96\sigma_1^2\Omega^{5/6}\right)^2}{\Omega^2\sqrt{\left(1+2.96\sigma_1^2\Omega^{5/6}\right)^2+1.2\sigma_1^2\Omega^{5/6}}} \right]$ \\
		$\langle \Delta x_0^2\rangle$,$\langle \Delta y_0^2\rangle$ & $0.33W_0^2\sigma_1^2\Omega^{-7/6}$ \\
		$\langle\Delta\Theta_{1,2}^2\rangle$ & $\ln\left[1+\frac{1.2\sigma_1^2\Omega^{5/6}}{\left(1+2.96\sigma_1^2\Omega^{5/6}\right)^2}\right]$ \\
		$\langle\Delta\Theta_1\Delta\Theta_2\rangle$ & $\ln\left[1-\frac{0.8\sigma_1^2\Omega^{5/6}}{\left(1+2.96\sigma_1^2\Omega^{5/6}\right)^2}\right]$ \\
		\bottomrule[0.75pt]
		& Strong Turbulence \\
		\midrule[0.5pt]
		$\langle \Theta_{1,2} \rangle$ & $\ln\left[\frac{\left(\gamma+1.71\sigma_1^{12/5}\Omega^{-1}-2.99\sigma_1^{8/5}\Omega^{-1}\right)^2}{\sqrt{\left(\gamma+1.71\sigma_1^{12/5}\Omega^{-1}-2.99\sigma_1^{8/5}\Omega^{-1}\right)^2+3.24\gamma\sigma_1^{12/5}\Omega^{-1}}}\right]$ \\
		$\langle \Delta x_0^2\rangle$,$\langle \Delta y_0^2\rangle$ & $0.75W_0^2\sigma_1^{8/5}\Omega^{-1}$ \\
		$\langle\Delta\Theta_{1,2}^2\rangle$ & $\ln\left[1+\frac{13.14\gamma\sigma_1^{12/5}\Omega^{-1}}{\left(\gamma+1.71\sigma_1^{12/5}\Omega^{-1}-2.99\sigma_1^{8/5}\Omega^{-1}\right)^2}\right]$ \\
		$\langle\Delta\Theta_1\Delta\Theta_2\rangle$ & $\ln\left[1+\frac{0.65\gamma\sigma_1^{12/5}\Omega^{-1}}{\left(\gamma+1.71\sigma_1^{12/5}\Omega^{-1}-2.99\sigma_1^{8/5}\Omega^{-1}\right)^2}\right]$ \\
		\bottomrule[0.75pt]
	\end{tabular}
\end{table}

\section{Scintillation Index for Weak Turbulence}
\label{scintillationIndexWeak}
The scintillation index can be expressed as \cite{Andrews2001}
\begin{equation}
\sigma_I^2(\textbf{r},L) = \sigma_{I,r}^2(\textbf{r},L) + \sigma_I^2(\textbf{0},L),
\label{SI}
\end{equation}
where $\sigma_{I,r}^2(\textbf{r},L)$ and $\sigma_I^2(\textbf{0},L)$ are radial and longitudinal component respectively.

Considering Kolmogorov spectrum, the radial component has a simple form
\begin{equation}
\begin{split}
\sigma_{I,r}^2(\textbf{r},L) &= 2.65\sigma_1^2 \Lambda^{5/6} \big[ 1 \\
&- \sideset{_1}{_1}{\mathop{F}}\left( -5/6;1;2r^2/W^2 \right) \big],
\end{split}
\label{WeakSIRadial}
\end{equation}
where $\sideset{_1}{_1}{\mathop{F}}(a;b;x)$ is the confluent hypergeometric function, and the longitudinal component is given as
\begin{equation}
\begin{split}
\sigma_I^2(\textbf{0},L) &= 3.86\sigma_1^2 {\rm Re}\Big[ -11/16\times\Lambda^{5/6} \\
&+i^{5/6} \sideset{_2}{_1}{\mathop{F}}\left( -5/6,11/6;17/6;\bar{\Theta}+i\Lambda \right)\Big],
\end{split}
\label{WeakSILon}
\end{equation}
where $\sideset{_2}{_1}{\mathop{F}}(a,b;c;x)$ is the hypergeometric function.

The beam parameters in Eq.(\ref{WeakSIRadial}) and (\ref{WeakSILon}) are defined by
\begin{equation}
\begin{split}
&\Theta = \frac{\Omega^2}{1+\Omega^2}, \; \Lambda = \frac{\Omega}{1+\Omega^2}\\ 
&\bar{\Theta} = 1 - \Theta, \; W = \frac{W_0}{\sqrt{1+\Omega^{-2}}} .
\end{split}
\end{equation}
There is also a approximate form
\begin{equation}
\begin{split}
&\sigma_I^2(\textbf{r},L) \approx 4.42\sigma_1^2\Lambda^{5/6}\frac{r^2}{W^2} + 3.86\sigma_1^2 \Bigg\{ -\frac{11}{16}\Lambda^{5/6} \\
&+0.4\big[ (1+2\Theta)^2 + 4\Lambda^2 \big]^{5/12} \cos\bigg[ \frac{5}{6}\tan^{-1}\left(\frac{1+2\Theta}{2\Lambda}\right) \bigg] \Bigg\}
\end{split}
\end{equation}
if needed. In this paper, Eq. (\ref{WeakSIRadial}) and (\ref{WeakSILon}) are applied in calculations.

\section{Scintillation Index for Strong Turbulence}
\label{scintillationIndexStrong}
The scintillation index still comprises radial and longitudinal component as indicated in Eq. (\ref{SI}).

The radial component can be expressed as
\begin{equation}
\label{eq30}
\sigma_{I,r}^2(\textbf{r},L) = 4.42\sigma_1^2\Lambda_e^{5/6}\frac{r^2}{W_e^2}, \quad r<W_e
\end{equation}
when the outer scale is not very large, where
\begin{equation}
W_e = W\sqrt{1+1.63\sigma_1^{12/5}\Lambda} ,\quad \Lambda_e = 2L/kW_e^2
\end{equation}
represent the effective beam parameters. However, the radial component in Eq.(\ref{eq30}) is quite sensitive to outer-scale effects when the outer scale is large enough, and it is given as
\begin{equation}
\sigma_{I,r}^2(\textbf{r},L) = 4.42\sigma_1^2\Lambda_e^{\frac{5}{6}} \left[ 1 - 1.15\left(\frac{\Lambda_eL}{kL_0^2}\right)^{\frac{1}{6}} \right] \frac{r^2}{W_e^2},
\end{equation}
where $L_0$ is the outer scale.

The longitudinal component is given by
\begin{equation}
\label{eq33}
\sigma_I^2(\textbf{0},L) = \exp\left( \sigma_{\ln x}^2 + \sigma_{\ln y}^2 \right) - 1,
\end{equation}
where $\sigma_{\ln x}^2$ and $\sigma_{\ln y}^2$ are large-scale and small-scale log-irradiance variances, respectively. Here exists the relations
\begin{align}
\alpha & = \left[ \exp(\sigma_{\ln x}^2) - 1 \right]^{-1} \\
\beta & = \left[ \exp(\sigma_{\ln y}^2) - 1 \right]^{-1}
\end{align}
where $\alpha$ and $\beta$ are the effective number of large scale and small scale cells in gamma-gamma distribution Eq.(\ref{eq25}), respectively. When inner scale and outer scale effects are both involved, the longitudinal component can be expressed as
\begin{equation}
\begin{split}
\sigma_I^2(\textbf{0},L) = & \exp\big[ \sigma_{\ln x}^2(l_0) - \sigma_{\ln x}^2(L_0) \\
& + \sigma_{\ln y}(l_0)^2 \big] - 1.
\end{split}
\end{equation}
where $\sigma_{\ln x}^2(l_0)$ with inner scale $l_0$ is given by
\begin{equation}
\begin{split}
&\sigma_{\ln x}^2(l_0) = 0.49\sigma_1^2\left( \frac{1}{3} - \frac{1}{2}\bar{\Theta} +\frac{1}{5}\bar{\Theta}^2 \right) \left( \frac{\eta_x Q_l}{\eta_x+Q_l} \right)^{\frac{7}{6}} \\
& \times\left[ 1 + 1.75\sqrt{\frac{\eta_x}{\eta_x+Q_l}} - 0.25\left( \frac{\eta_x}{\eta_x+Q_l} \right)^\frac{7}{12} \right]
\end{split}
\end{equation}
and
\begin{equation}
\begin{split}
\frac{1}{\eta_x} = & \frac{0.38}{1-3.21\bar{\Theta}+5.29\bar{\Theta}^2} \\
& + 0.47\sigma_1^2Q_l^{\frac{1}{6}} \left( \frac{\frac{1}{3} - \frac{1}{2}\bar{\Theta} +\frac{1}{5}\bar{\Theta}^2}{1+2.2\bar{\Theta}} \right)^{\frac{6}{7}},
\end{split}
\end{equation}
where $Q_l=10.89L/kl_0^2$. Similar to $\sigma_{\ln x}^2(l_0)$, the $\sigma_{\ln x}^2(L_0)$ is given as
\begin{equation}
\begin{split}
&\sigma_{\ln x}^2(L_0) = 0.49\sigma_1^2\left( \frac{1}{3} - \frac{1}{2}\bar{\Theta} +\frac{1}{5}\bar{\Theta}^2 \right) \left( \frac{\eta_{x0} Q_l}{\eta_{x0}+Q_l} \right)^{\frac{7}{6}} \\
& \times\left[ 1 + 1.75\sqrt{\frac{\eta_{x0}}{\eta_{x0}+Q_l}} - 0.25\left( \frac{\eta_{x0}}{\eta_{x0}+Q_l} \right)^\frac{7}{12} \right]
\end{split}
\end{equation}
where $\eta_{x0} = \eta_x Q_0/(\eta_x + Q_0)$, and $Q_0 = 64\pi^2L/kL_0^2$ is a nondimensional outer-scale parameter. The small-scale log-irradiance variance $\sigma_{\ln y}^2(l_0)$ can be written as
\begin{equation}
\sigma_{\ln y}^2(l_0) = \frac{0.51\sigma_G^2}{\left(1 + 0.69\sigma_G^{12/5}\right)^{5/6}}
\end{equation}
where $\sigma_G^2$ is the weak fluctuation scintillation index and can be written as
\begin{equation}
\begin{split}
&\sigma_G^2 = 3.86\sigma_1^2\Bigg\{ 0.4\frac{[(1+2\Theta)^2+(2\Lambda+3/Q_l)^2]^{\frac{11}{12}}}{\sqrt{(1+2\Theta)^2+4\Lambda^2}} \\
& \Bigg[ \frac{2.61}{[(1+2\Theta)^2Q_l^2+(3+2\Lambda Q_l)^2]^{\frac{1}{4}}}\sin\left( \frac{4}{3}\varphi_2+\varphi_1 \right) \\
& -\frac{0.52}{[(1+2\Theta)^2Q_l^2+(3+2\Lambda Q_l)^2]^{\frac{7}{24}}}\sin\left( \frac{5}{4}\varphi_2+\varphi_1 \right) \\
& +\sin\left( \frac{11}{6}\varphi_2+\varphi_1 \right) \Bigg] - \frac{13.4\Lambda}{Q_l^{\frac{11}{6}}[(1+2\Theta)^2+4\Lambda^2]} \\
& -\frac{11}{6}\Bigg[ \left( \frac{1+0.31\Lambda Q_l}{Q_l} \right)^{\frac{5}{6}} + \frac{1.1(1+0.27\Lambda Q_l)^{\frac{1}{3}}}{Q_l^{\frac{5}{6}}} \\
& - \frac{0.19(1+0.24\Lambda Q_l)^{\frac{1}{4}}}{Q_l^{\frac{5}{6}}} \Bigg] \Bigg\},
\end{split}
\end{equation}
and
\begin{align}
\label{eq44}
\varphi_1 & = \tan^{-1}\left( \frac{2\Lambda}{1+2\Theta} \right), \\
\varphi_2 & = \tan^{-1}\left[ \frac{(1+2\Theta)Q_l}{3+2\Lambda Q_l} \right].
\end{align}

\section{the Mean Irradiance}
\label{MeanIrradiance}
The (normalized) mean irradiance can be approximated by the Gaussian function \cite{Andrews2001}
\begin{equation}
\label{eq26}
\langle I(\textbf{r},L)\rangle = \frac{W_0^2}{W_e^2}\exp\left( -2r^2/W_e^2\right),
\end{equation}
where $r$ is the distance from the beam center line in the transverse direction, and $W_e$ is a measure of the effective beam spot size given by
\begin{equation}
W_e =
\begin{cases}
W\sqrt{ 1 + 1.33\sigma_1^2\Lambda^{\frac{5}{6}}},\: {\rm weak\:fluctuations} \\
W\sqrt{ 1 + 1.63\sigma_1^{\frac{12}{5}}\Lambda},\; {\rm strong\:fluctuations}
\end{cases}.
\end{equation}

\end{document}